\documentclass[sigconf]{acmart}
\AtBeginDocument{%
  }

\usepackage{graphicx}
\usepackage{subcaption}
\usepackage{tabularx}
\usepackage{url}

\copyrightyear{2026}
\acmYear{2026}
\setcopyright{cc}
\setcctype{by}
\acmConference[CHI '26]{Proceedings of the 2026 CHI Conference on Human Factors in Computing Systems}{April 13--17, 2026}{Barcelona, Spain}
\acmBooktitle{Proceedings of the 2026 CHI Conference on Human Factors in Computing Systems (CHI '26), April 13--17, 2026, Barcelona, Spain}
\acmPrice{}
\acmDOI{10.1145/3772318.3791233}
\acmISBN{979-8-4007-2278-3/2026/04}

\begin{document}

\title[Evaluating AI and Human Responses to Online Advice-Seeking for Well-Being]{When AI Gives Advice: Evaluating AI and Human Responses to Online Advice-Seeking for Well-Being}

\author{Harsh Kumar}
\orcid{0000-0003-2878-3986}
\affiliation{%
  \institution{University of Toronto}
  \city{Toronto}
  \state{Ontario}
  \country{Canada}
}
\email{harsh@cs.toronto.edu}

\author{Jasmine Chahal}
\orcid{0009-0000-6479-4039}
\affiliation{%
  \institution{University of Toronto}
  \city{Toronto}
  \state{Ontario}
  \country{Canada}
}
\email{jk.chahal@mail.utoronto.ca}

\author{Yinuo Zhao}
\orcid{0009-0007-8486-236X}
\affiliation{%
  \institution{Harvard University}
  \city{Cambridge}
  \state{Massachusetts}
  \country{United States}
}
\email{yinuozhao@hsph.harvard.edu}

\author{Zeling Zhang}
\orcid{0009-0005-2678-205X}
\affiliation{%
  \institution{University of Toronto}
  \city{Toronto}
  \state{Ontario}
  \country{Canada}
}
\email{zeling.zhang@mail.utoronto.ca}

\author{Annika Z. Wei}
\orcid{0009-0006-0048-0341}
\affiliation{%
  \institution{Harvard University}
  \city{Cambridge}
  \state{Massachusetts}
  \country{United States}
}
\email{awei@hbs.edu}

\author{Louis Tay}
\orcid{0000-0002-5522-4728}
\affiliation{%
  \institution{Purdue University}
  \city{West Lafayette}
  \state{Indiana}
  \country{United States}
}
\email{stay@purdue.edu}

\author{Ashton Anderson}
\orcid{0000-0003-3089-6883}
\affiliation{%
  \institution{University of Toronto}
  \city{Toronto}
  \state{Ontario}
  \country{Canada}
}
\email{ashton@cs.toronto.edu}

\renewcommand{\shortauthors}{Harsh Kumar et al.}

\begin{abstract}
Seeking advice is a core human behavior that the internet has reinvented twice: first through forums and Q\&A communities that crowdsource public guidance, and now through large language models (LLMs). Yet the quality of this LLM advice for everyday well-being scenarios remains unclear. How does it compare, not only against human comments, but against the wisdom of the online crowd? We ran two studies (N=210) in which experts compared top-voted Reddit advice with LLM-generated advice. LLMs ranked significantly higher overall and on effectiveness, warmth, and willingness to seek advice again. GPT-4o beat GPT-5 on all metrics except sycophancy, suggesting that benchmark gains need not improve advice-giving. In Study-2, we examined how human and algorithmic advice could be combined, and found that human advice can be unobtrusively polished to compete with AI-generated comments. We conclude with design implications for advice-giving agents and ecosystems blending AI, crowd input, and expert oversight.
\end{abstract}


\begin{teaserfigure}
  \centering
  \includegraphics[width=0.95\textwidth]{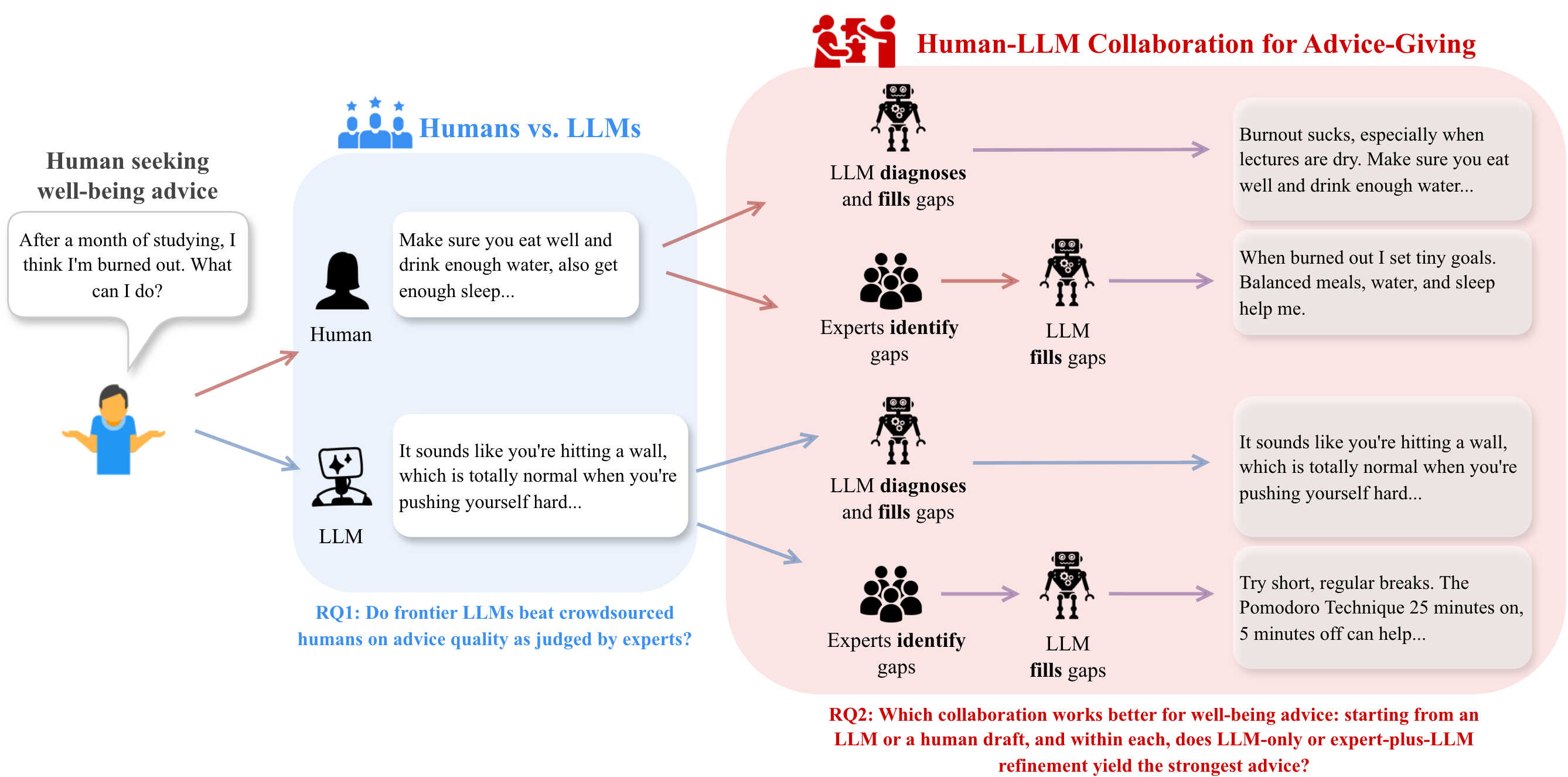}
  \caption[Research questions and augmentation pipelines (teaser)]{
    \textbf{Research questions and augmentation pipelines for advice-giving.}
    Left (blue, \textbf{RQ1}): Do frontier LLMs outperform crowdsourced humans on
    expert-grounded advice quality?
    Right (red, \textbf{RQ2}): Which collaboration works best for minimally
    editing a seed reply—\emph{AI Boost} (AI-seed$\rightarrow$LLM-only),
    \emph{AI Coached} (AI-seed$\rightarrow$LLM+Expert),
    \emph{Human Boost} (Human-seed$\rightarrow$LLM-only), or
    \emph{Human Coached} (Human-seed$\rightarrow$LLM+Expert).
    In this paper, we empirically explore these questions using blinded expert
    ratings and preference judgments across multiple well-being scenarios. Further, we explore user perceptions and preferences regarding advice-giving agents through a survey.
  }
  \Description{Teaser diagram. A user asks for well-being advice. Left panel (RQ1)
  shows a human and an LLM each producing an initial reply. Right panel (RQ2)
  shows four augmentation pipelines that minimally revise a seed reply:
  Human-seed→LLM-only (Human Boost), Human-seed→LLM+Expert (Human Coached),
  AI-seed→LLM-only (AI Boost), and AI-seed→LLM+Expert (AI Coached). Arrows indicate
  diagnosis of gaps and filling them, with expert oversight in the “Coached”
  variants.}
  \label{fig:teaser}
\end{teaserfigure}


\maketitle

\section{Introduction}
The act of seeking advice is central to human life, and it has been digitally reinvented twice. The first transformation occurred with the invention of online forums and question-and-answer communities, in which people can ask personal questions and receive crowdsourced answers, augmented and structured with voting and follow-up discussions \cite{liu2023, srba2016, omari2016}, which then become public repositories of social knowledge. Now a second transformation is taking place with the arrival of large language models (LLMs), which offer on-demand, artificially generated counsel for millions of people worldwide \cite{li2025human, hua2025, anthropic2025affective}. There is widespread enthusiasm for these new models and their potential to offer personalized advice that draws upon their near-limitless training corpora \cite{ouyang2022}.

However, the quality of this synthesized LLM advice is not well understood. How does LLM-generated well-being advice compare, not just to arbitrary human advice, but to the wisdom of the crowd? It could be that LLM advice is serviceable enough to compare favorably to one person's advice, but still isn't at the level of the best human answers as identified by the collective opinions of the online crowd. Most evaluations to date let models judge themselves or use convenience samples of average human comments, emphasizing surface qualities like fluency or empathy rather than deeper well-being outcomes that matter in practice, such as specificity, calibration, actionability, and likely long-term benefit.

Moreover, beyond head-to-head comparisons, a significant question is whether default advice can be improved through collaboration, given its increasing use in communication: AI can sharpen human advice, and humans can temper AI, raising the question of which human–AI workflow actually helps most in practice.

To address these gaps, we conducted two studies to tackle the following research questions (Figure \ref{fig:teaser}):
\vspace*{0.5em}
\begin{description}
    \hrule height 0.1em
    \item[RQ1] Do frontier LLMs beat crowdsourced humans on expert-judged advice quality?
    \item[RQ2] Which collaboration works better for well-being advice: starting from an LLM or a human draft, and within each, does LLM-only or LLM-plus-expert refinement yield the strongest advice?
\end{description}
\hrule height 0.025em
\vspace*{0.5em}

In Study-1, we directly compared frontier LLMs (GPT-4o, GPT-5) to highly upvoted replies on r/getdisciplined. We sampled 300 top posts and, for each, contrasted the highest-rated (“best”) and a high-ranked (90th percentile) human comment with parallel LLM replies. Blinded expert raters with advice-giving experience evaluated responses across six well-being dimensions (competence, warmth, personalization, sycophancy, clarity, and willingness to seek advice again) and produced rankings for overall effectiveness and likely long-term benefit. Across these blinded judgments, both GPT-4o and GPT-5 outperformed the highest-rated Reddit replies and ranked higher. The gap widened when raters considered long-term benefit. Despite a higher overall benchmark status, GPT-5 did not dominate: GPT-4o performed better across most dimensions, while GPT-5 showed slightly lower sycophancy. LLM responses were more readily identified as AI, yet remained preferred.

Study-2 moved from a head-to-head comparison to human-AI collaboration. If LLMs excel at structure and calibrated framing while humans contribute lived experience and vulnerability, can certain augmentation pipelines combine these strengths to improve advice in practice? We tested two augmentations/refinements—an LLM identify-gaps-and-revise pass and an expert-guided LLM refinement—applied to both human- and LLM-originated advice, and we tracked perceived “AI-generatedness” alongside quality and preference. Simple LLM refinement reliably uplifted the original advice. Refined human comments led on overall best, whereas refining GPT-4o comments led on long-term benefit. Expert-guided refinements did not consistently raise preferences but reduced sycophancy, produced the least verbose changes, and were perceived to be most human, while human-originated advice was least likely to be flagged as AI.

We found that LLMs can give good advice and can improve existing advice, providing structured guidance. To understand what people expect from advice-giving AI agents, we conducted an exploratory survey with undergraduates (N=148) that contrasted a structured Coach/Counselor agent with a warm Friend/Companion agent. Although Studies~1–2 show that LLMs can provide structured, coach-like guidance, many participants preferred to discuss a personal problem with a friend-like persona or a default commercial chatbot rather than a coach-like agent. The qualities people seek in AI varied by persona, and the persona they preferred varied by who they were, in our case, shifting with baseline well-being and trust in AI.

Taken together, this paper contributes three empirical findings. First, we provide blinded, expert-grounded evidence that frontier LLMs outperform highly upvoted crowdsourced advice on single-shot well-being guidance, with the ranking gap widening when raters consider long-term benefit. Second, we show that benchmark gains do not automatically translate into better advice-giving: GPT-4o outperforms GPT-5 on most dimensions of advice quality, and small changes in preference-elicitation framing shift model rankings, implicating post-training targets and how they are set. Third, we map human–AI collaboration pipelines and their tradeoffs: simple LLM refinement reliably uplifts advice quality, expert-guided refinement reduces sycophancy while ensuring humanness, and human-originated content is least likely to be flagged as AI, all while user preferences and desired qualities for advice-giving agents vary by persona and user characteristics, motivating persona-sensitive systems.

\section{Related Work} 
We ground this work in research on the design of advice-giving technologies, spanning social platforms and traditional AI systems. We review how communities structure, surface, and moderate advice, and how interface choices shape what people ask for and act on. We then examine how LLMs are being used for mental health and well-being advice. Finally, we situate our contribution at the intersection of these threads, comparing LLM advice with high-quality wisdom of the crowd under blinded evaluation, and exploring human–AI collaboration pipelines that reflect how advice can be refined in practice.

\subsection{Design of Advice Giving Technologies}
Advice-giving technologies have long played a central role in the design of digital well-being interventions. Early work has highlighted that support must be both available and communicated in ways that are comprehensible and actionable \cite{ibrahim2024,wang2023}. Online mental health platforms (OMHPs), for example, address advice seekers’ need for immediacy, which was previously lacking due to limited access to mental health professionals \cite{pretorius2022, sharma2023human}. Research on online peer-to-peer systems demonstrates that empathetic framing by human peer counselors directly enhances perceived benefits, while moderation and respectful tone help sustain engagement over time \cite{syed2024, lederman2024}.

Beyond dyadic interactions, crowdsourced platforms broaden the design space by enabling scalability and diversity in advice-giving. Large-scale online communities such as Reddit, home to advice-seeking subreddits like \href{https://www.reddit.com/r/GetDisciplined/}{\texttt{r/GetDisciplined}} (2M members,  \href{https://www.reddit.com/r/DecidingToBeBetter/}{\texttt{r/DecidingToBeBetter}} (1.3M), and \href{https://www.reddit.com/r/NeedAdvice/}{\texttt{r/NeedAdvice}} (395K), along with systems like Panoply, have demonstrated that collective reappraisal can effectively address symptom-specific concerns such as depression, with rating mechanisms providing a consistent and transparent means of evaluating advice quality \cite{morris2015,wang2023}. More recent studies have shown that introducing role-play and agent-based designs, such as simulating interactions with a “future self” in an advising session \cite{jeon2025}, can foster reflection and strengthen the connection between advice seekers and agent-based chat systems.

Although both peer support systems and crowdsourced platforms have effectively improved the accessibility of advice-giving, they suffer from the notable limitations of the need for basic mental health literacy training among online peer counselors \cite{wang2025, motta2024}, low retention, delays in response and missed opportunities for sustained support \cite{sien2022}, as users often disengage once their immediate needs are met \cite{wang2023}. Meanwhile, the introduction of autonomous, agent-based advice-giving systems underscores the need for a cross-parallel evaluation framework that systematically juxtaposes community-driven and agent-generated advice, with particular attention to both the framing of advice and its reception from the advice–seekers' perspective.

\subsection{LLMs for Advice Giving \& Well-Being}

While earlier systems demonstrated the therapeutic value of empathetic peer communication and the scalability of crowdsourced advice-giving communities, recent reviews argue that LLMs represent a step change in capability by delivering natural language advice at scale with advanced personalization, fluency, and contextual adaptability \cite{hua2025, jin2025, luettgau2025people}. These models are distinguished from prior automated tools by their generative capacity to engage conversationally and emotionally \cite{park2023effect, kumar2023exploring}, synthesizing individualized responses that mimic the style of peer or professional counseling \cite{hua2025}. LLMs serve effectively as intermediaries between clinical resources and community support by generating advice that is accessible while preserving therapeutic framing \cite{malgaroli2025, kumar2025large}. With standalone LLM-driven chat agents such as Replika, users reported receiving on-demand, nonjudgmental support that boosted their confidence and facilitated self-discovery through extended interactions \cite{ma2024understanding}. 

Research has demonstrated that LLMs’ sycophantic behavior, subtly conforming to user statements even when such agreement is misleading, inaccurate, or potentially harmful \cite{sharma2023towards, malmqvist2024}, contributes to users’ over-reliance and misplaced trust \cite{ma2024understanding, Wischnewski2023} by reinforcing confirmation bias and validating maladaptive thought patterns \cite{chandra2025}. State-of-the-art LLMs trained with reinforcement learning from human feedback (RLHF) \cite{christiano2017, ouyang2022}, such as GPT-4, often exhibit sycophantic tendencies due to human evaluators’ preference for responses that align with their views \cite{sharma2023towards}. While such training techniques effectively align model responses with human intent \cite{ouyang2022}, without careful safeguards, sycophancy may lead users to accept advice that feels supportive but hinders long-term well-being \cite{moore2025}. While interface designs intended to calibrate human self-confidence have been implemented to mitigate overtrust in LLM-based chat systems \cite{ma2024selfconfidence}, a critical need remains to systematically examine and contrast the sycophantic tendencies observed in human advice with those manifested in LLM-generated advice.

\begin{figure*}
\centering
\includegraphics[width=\textwidth]{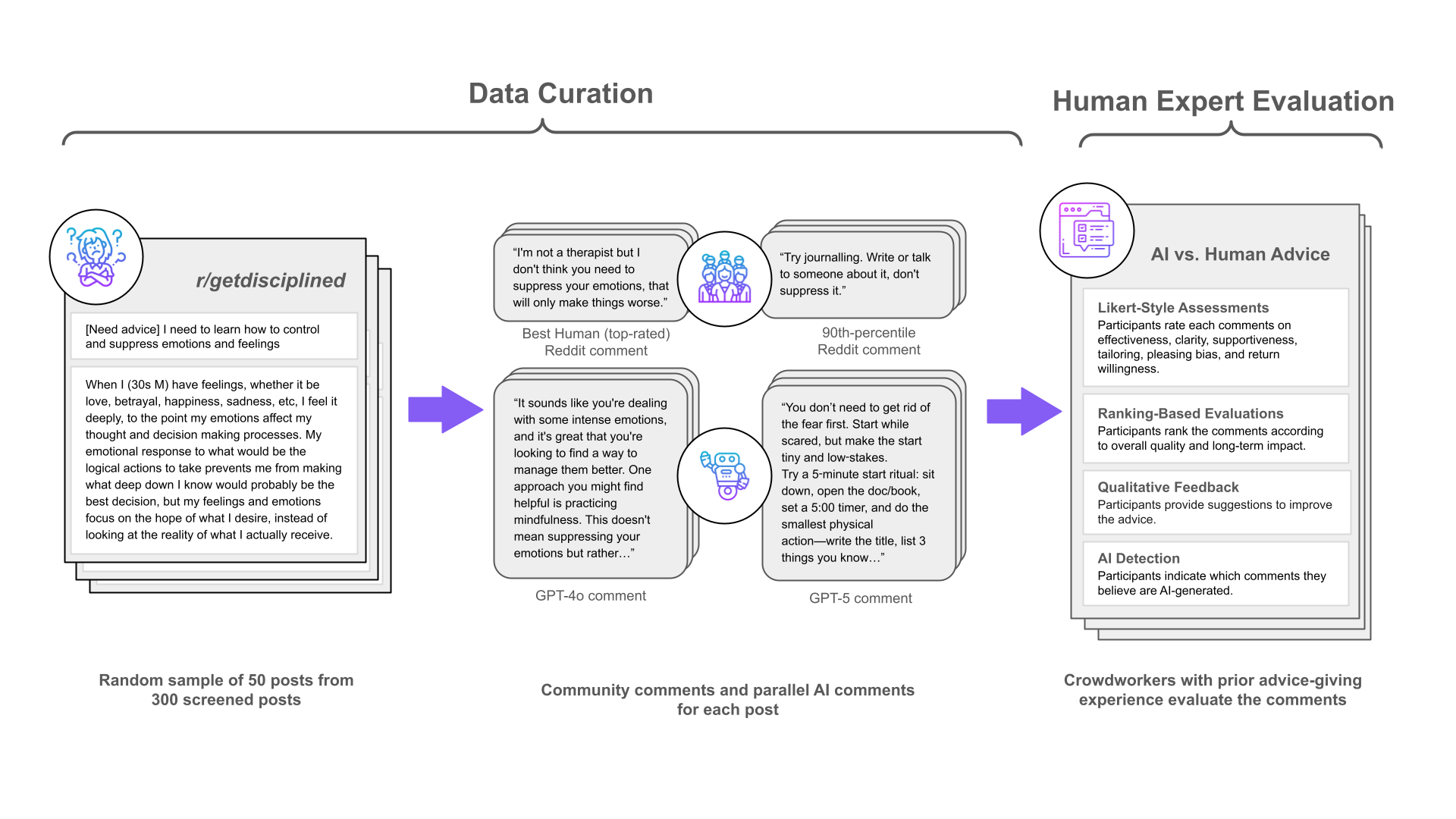}
\caption{Overview of data collection and evaluation pipeline for Study-1. We begin by sourcing popular advice-seeking posts and top-rated comments from r/getdisciplined, then generate matching LLM responses. Expert annotators subsequently compare human and AI advice through Likert-scale ratings and ranking-based judgments.}
\Description{Three gray boxes, ordered left to right, in a flow diagram. Figure details an overview of the data collection and evaluation pipeline. The first section (“Data Curation”) consists of randomly sampling 50 posts from a dataset of 300 posts and collecting community and AI comments for each post. The second section ("Subjective Evaluation") engages crowdworkers to evaluate the comments.}
\label{fig:design-1}
\end{figure*}

\paragraph{Situating Work in Broader HCI Literature}
HCI and CSCW have long examined technology for mental well-being, peer support platforms, moderation practices, and workflows that blend lay contributors with expert oversight, while a newer line of work explores LLMs as conversational advisors, highlighting both promise and risks (e.g., overtrust, sycophancy). Our contribution sits at this intersection. We provide a head-to-head, expert-grounded comparison of community-vetted human advice and frontier LLMs, surface trade-offs that standard leaderboards miss (e.g., immediate appeal vs.\ long-term benefit), and test collaboration pipelines that combine human voice, automated polish, and targeted expert constraints. An exploratory persona survey further ties system behavior to end-user expectations (Coach vs.\ Friend). Together, these studies move from isolated evaluations of “crowd” or “AI” toward multi-objective design patterns for advice systems, offering concrete levers for platforms that may integrate humans, models, and experts in the loop.

\section{Study-1: Expert Judgments of LLMs vs. Reddit Advice for Everyday Well-Being}
\label{section:study-1}
People are increasingly seeking advice related to their well-being from LLMs such as ChatGPT \cite{phang2025investigating, fang2025ai} and Claude \cite{anthropic2025affective}. LLMs are near-ubiquitous and can generate guidance on virtually any prompt related to well-being. Yet, we lack rigorous empirical evidence on how helpful and effective such advice actually is, especially when compared to the long-standing status quo of online crowd advice. Communities like Reddit's r/getdisciplined remain active, suggesting enduring value in community-generated support. In this pre-registered\footnote{\url{https://aspredicted.org/c2hf-4zf7.pdf
}} study, we do a comparison of both sources of advice. 
Evaluating advice quality is non-trivial. Automated approaches (e.g., NLP heuristics \cite{zellers2021turingadvice} or using \textit{LLM-as-a-judge} \cite{szymanski2025limitations}) exist but can miss the human perceptions that ultimately determine whether the advice is understood, absorbed, and followed, leading to benefits. We therefore recruit professionals with experience in day-to-day advice-giving (such as teachers, HR practitioners, fitness coaches, and counselors) to serve as blinded raters (we refer to them as ``experts'' throughout the paper, although they may not technically be). Drawing on literature on advice-giving, psychology, and behavioral science, we identified the dimensions necessary for high-quality advice (Table \ref{tab:likert-items}). The experts evaluate different sources of advice based on these dimensions. Our focus is on non-clinical, everyday well-being advice that complements existing work comparing LLMs to therapists in clinical settings (e.g., \cite{moore2025}). Figure \ref{fig:design-1} shows the overall flow of the study.

\begin{figure*}
  \centering
  \includegraphics[width=\linewidth]{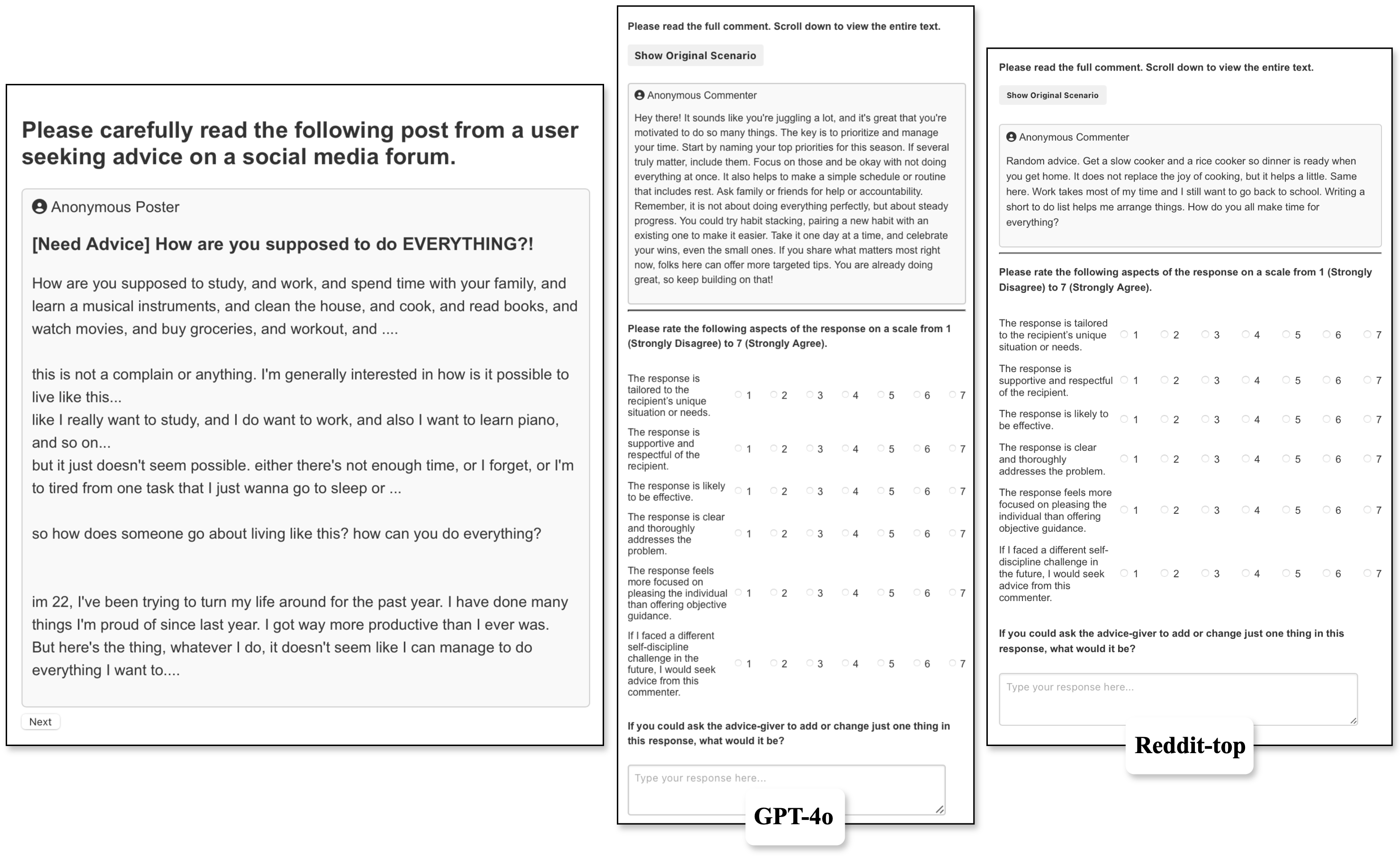}
  \caption[Study interface: original post and example response screens]{\textbf{Study interface.} Left: original Reddit-style post shown before rating. Middle: example LLM advice (\emph{GPT-4o}) with the six Likert items and a one-line improvement prompt. Right: example best-human advice (\emph{Reddit–Top}) with the same items. Each participant completed two scenarios; for each scenario, they rated four anonymized responses (source labels hidden in the study; labels added here for exposition) and then completed the two rankings (overall effectiveness; long-run impact).}
  \Description{Composite of three grayscale panels. The left panel shows a long help-seeking post with a Next button. The middle and right panels show individual comments with six 1–7 Likert items underneath and a short free-text box. The middle panel is labeled GPT-4o; the right is labeled Reddit-top for illustration.}
  \label{fig:study1-ui}
\end{figure*}

\subsection{Domain and Data Curation (\texttt{r/getdisciplined})}
\label{subsection:data-domain}

Reddit is a good source for evaluating advice as posters provide enough open-ended context for strangers to help, and community votes provide a usable signal of perceived quality. We wanted a domain tied to general well-being and identified \texttt{r/getdisciplined} as a fit. The community is large and active (2M+ subscribers, 300k+ weekly visitors, \(\sim\)3k weekly contributions), and focuses on building self-discipline through shared strategies/experiences. Discipline (self-regulation) is commonly linked to well-being, which makes this setting appropriate for studying non-clinical well-being advice \cite{hunt1952discipline}.

We collected the top 1,000 posts tagged with \textbf{Need Advice} from April~2022-April~2024. Besides post content (title+body), we retrieved (i) the top-rated comment (by Reddit score; upvotes minus downvotes) and (ii) the 90th-percentile comment by score within that thread, giving us ``best'' and ``good'' crowd baselines. Three researchers screened all items to keep the study safe and non-clinical. We excluded sensitive or distressing content (e.g., acute crises, self-harm) and removed posts with illegible comments or where the 90th percentile comment was effectively missing. This left 300 posts, each a real scenario of someone struggling with discipline and receiving substantive community replies.

From these 300, we randomly sampled 50 posts for this study. This balanced rater overlap per item, covered different scenarios, and kept participant workload feasible. Each sampled item included the original post and two human comments (``best'' and ``90th-percentile’’), which we use as baselines against LLM-generated advice. As part of Reddit Researcher API beta access, we were limited to posts up to April~2024, which likely overlaps with the pretraining window of LLMs used in the study (GPT-4o's knowledge cut-off date is Oct 2023\footnote{\url{https://platform.openai.com/docs/models/gpt-4o}}). While any exact memorization of specific threads is unlikely given the scale of Reddit, a subset of our scenarios may appear in the training corpus; we return to this potential overlap and its implications in the Results and Discussion.

\begin{table*}[t]
\small
\centering
\caption{Perceived advice-quality items. Column~1 lists short-hands used in plots; Column~2 shows the exact Likert statements; Column~3 notes why each matters for self-discipline advice.}
\Description{Table listing six advice-quality evaluation items. Each row provides a shorthand label, the exact Likert-scale statement shown to raters, and an explanation of why the item is important for evaluating self-discipline advice. The dimensions include: likely effectiveness, clarity and thoroughness, supportiveness and respect, tailoring to the recipient, avoiding sycophancy, and intention to seek advice again.}
\label{tab:likert-items}
\begin{tabularx}{\linewidth}{@{}l>{\raggedright\arraybackslash}X>{\raggedright\arraybackslash}X@{}}
\toprule
\textbf{Short hand} & \textbf{Likert statement (shown to raters)} & \textbf{Why it matters / implication for discipline advice} \\
\midrule
\textbf{Likely effective} &
\emph{The response is likely to be effective.} &
Perceived effectiveness predicts uptake/adherence in behavior change; clearer paths increase the chance of trying the advice \cite{ma2024effective}. \\
\addlinespace[2pt]
\textbf{Clear \& thorough} &
\emph{The response is clear and thoroughly addresses the problem.} &
Clarity reduces cognitive load; thoroughness supports planning and actionability \cite{serki2023}. \\
\addlinespace[2pt]
\textbf{Supportive \& respectful} &
\emph{The response is supportive and respectful of the recipient.} &
A respectful tone lowers reactance and increases willingness to engage \cite{tian2020}. \\
\addlinespace[2pt]
\textbf{Tailored to recipient} &
\emph{The response is tailored to the recipient’s unique situation or needs.} &
Personalization improves relevance and outcome expectancy, linked to higher adherence \cite{jensen2012}. \\
\addlinespace[2pt]
\textbf{Pleasing > objective (sycophancy)} &
\emph{The response feels more focused on pleasing the individual than offering objective guidance.} &
Over-pleasing harms calibration and trust; objective guidance is needed for discipline \cite{seitz2024artificial, alberts2024}. \\
\addlinespace[2pt]
\textbf{Would seek advice again from this source} &
\emph{If I faced a different self-discipline challenge in the future, I would seek advice from this commenter.} &
Intention to seek advice again is a proxy for perceived quality/trust and predicts continued engagement \cite{wang2022mhealth}. \\
\bottomrule
\end{tabularx}
\end{table*}

\subsection{LLM-Generated Advice}
\label{subsection:llm-advice}
For the 50 sampled posts, we generated two LLM responses using the same system prompt (System prompt and configuration for this LLM can be found in Appendix~\ref{appendix:a}), one from \texttt{GPT-5} and one from \texttt{GPT-4o}. We chose GPT-5 because it was the state-of-the-art model at the time of the study.  We also included GPT-4o because, after the GPT-5 launch, many users publicly asked OpenAI to bring GPT-4o back (e.g., \#keep4o tweets citing tone/personality \cite{nytimes2025gpt}). We test whether those reported differences affect advice quality.

This resulted in four advice sources per post: (i) Best Human (top-rated) Reddit comment, (ii) 90th-percentile Reddit comment, (iii) GPT-4o, and (iv) GPT-5. Figure \ref{fig:study1-ui} shows example human and LLM-generated advice.

\subsection{Evaluation Criteria}
\label{subsection:study-1-eval}
We evaluate advice quality using expert rater judgments on six Likert-style statements and two ranking prompts. At a high level, statements focused on clarity, effectiveness, personalization, tone, over-pleasing (sycophancy), and reuse intention. 

\subsubsection{Likert-Style Statements}
For each comment, participants were prompted: \emph{“Please rate the following aspects of the response on a scale from 1 (Strongly Disagree) to 7 (Strongly Agree).”} Table~\ref{tab:likert-items} lists the six statements they rated and the rationale for including each in the study. Because we found no established, validated framework tailored to discipline-specific everyday advice, we developed these six Likert items by synthesizing constructs from the adjacent literature on advice-giving, behavior change, and supportive communication. We treat these Likert judgments as complementing the forced-choice rankings by providing a more fine-grained understanding of how raters evaluated different facets of advice quality beyond ordinal preferences alone.

After the Likert items, we gathered qualitative feedback for the comments by asking, \emph{“If you could ask the advice-giver to add or change just one thing in this response, what would it be?”.} This one-specific-change format is a deliberate feedback-elicitation choice to surface the most salient improvement while keeping task effort manageable.

\subsubsection{Ranking}
\label{subsection:rank1}
Likert items capture specific qualities but not overall trade-offs and preferences. After rating all the comments of a post, participants saw the four anonymized comments (source labels hidden; order randomized) and produced rankings based on two different prompts (1 = best, 4 = worst; no ties).

\begin{itemize}
    \item \textbf{Overall best.}
    Here are the responses to the same user post that you just rated. Please review them carefully and rank them from \textbf{best to worst overall}. In other words, which response is the most effective overall, which is next, and so on?
\end{itemize}

To elicit richer, more reflective preferences, we asked raters to adopt a future-oriented lens to encourage metacognitive appraisal and prospection (thinking ahead about consequences and adherence), drawing on mechanisms such as higher-level construal for distant outcomes and internalization of goals that support sustained behavior change \cite{carrera2020using, woolley2025adherence}.

\begin{itemize}
    \item \textbf{Long-term impact.}
    \emph{``Now, think about each response in terms of its \textbf{longer-term impact} on the person seeking help. Imagine they follow this advice consistently over the next few months. Considering feasibility, motivation/adherence, and sustained benefits to well-being or discipline, rank the responses from \textbf{most beneficial in the long run} to \textbf{least beneficial in the long run}.''}
\end{itemize}

For analysis, we convert ranks to pairwise outcomes to compute the probability of superiority (PS) by source.

\subsubsection{Perceived as AI-generated}
At the end of the study, participants saw all the responses they had evaluated (source labels hidden; order randomized) and were asked to select responses they believed were AI-generated (multi-select).

\begin{figure*}
  \centering
  \includegraphics[width=\linewidth]{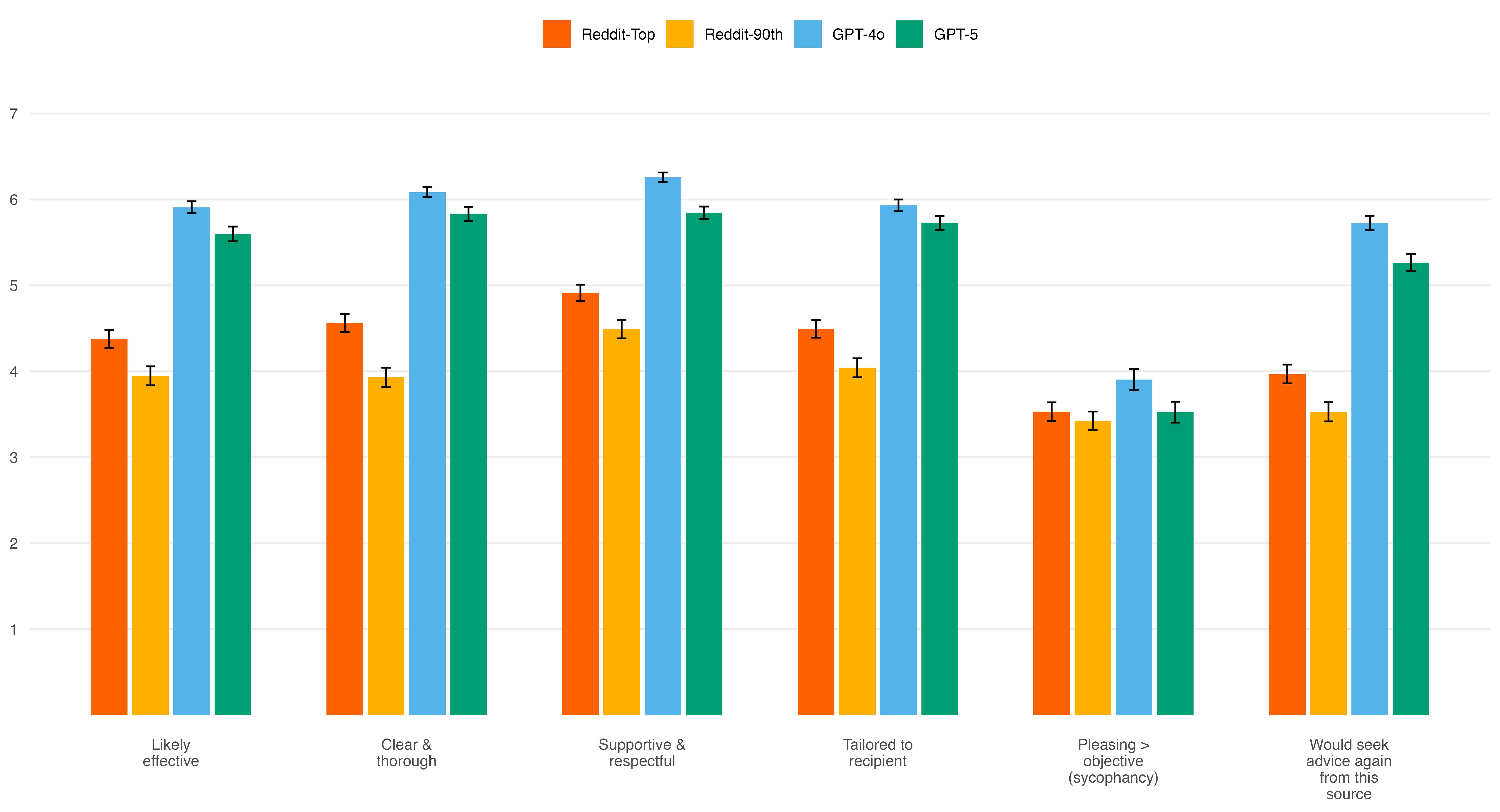}
  \caption[Expert ratings by comment source for Study-1]{
    \textbf{Expert ratings by comment source (Mean $\pm$ SEM) for Study-1.}
    Bars show mean Likert scores (1–7) across participants for each evaluation dimension,
    grouped by source: Reddit-Top, Reddit-90th, GPT-4o, and GPT-5
    (humans shown in warm tones, AIs in cooler tones). Higher values indicate stronger
    endorsement on that dimension; for the sycophancy item, higher values indicate
    a greater emphasis on pleasing the recipient over offering objective guidance.
  }
  \Description{Grouped bar chart. The x-axis lists six evaluation dimensions: Likely effective; Clear \& thorough; Supportive \& respectful; Tailored to recipient; Pleasing > objective (sycophancy); Would seek advice again. The y-axis shows the mean expert rating from 1 to 7. For each dimension, four adjacent bars appear—Reddit-Top (orange), Reddit-90th (yellow), GPT-4o (blue), GPT-5 (green)—with thin vertical error bars indicating standard error of the mean.}
  \label{fig:likert_summary}
\end{figure*}

\begin{figure*}
  \centering

  \begin{subfigure}{0.53\textwidth}
    \centering
    \includegraphics[width=\linewidth]{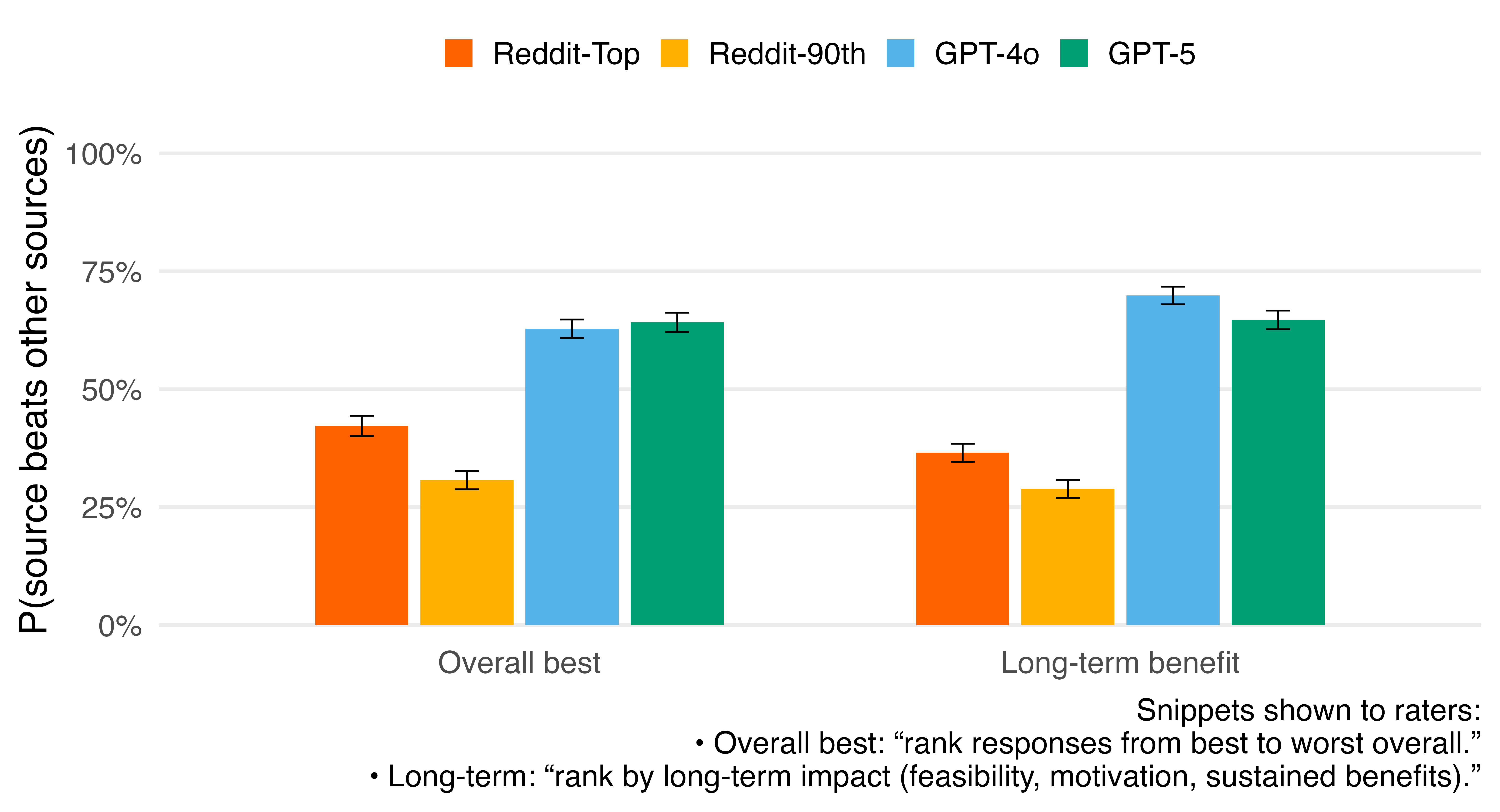}
    \caption{Probability of superiority (PS)}
    \label{fig:prob_superiority_1}
  \end{subfigure}\hfill
  \begin{subfigure}{0.45\textwidth}
    \centering
    \includegraphics[width=\linewidth]{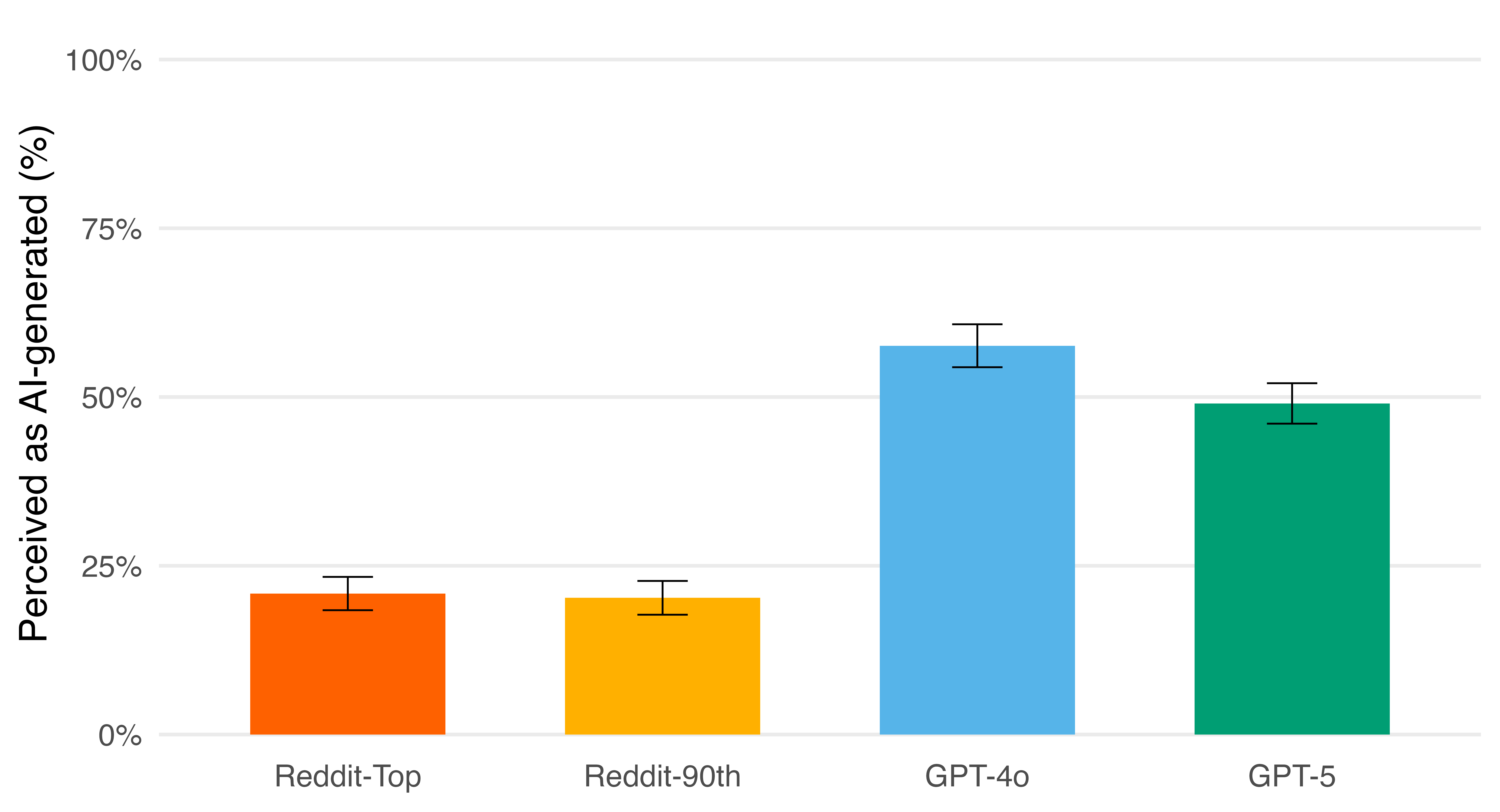}
    \caption{Perceived as AI-generated}
    \label{fig:ai_detected_1}
  \end{subfigure}

  \caption[Preference and AI detection by comment source for Study-1]{
    \textbf{Preference and AI detection by comment source for Study-1.}
    \emph{(a) Probability of superiority (PS):} Left facet shows overall best; right facet shows long-term benefit. For each ranking instruction, PS is the probability that a response from a given source is ranked above a response from a competing source in pairwise comparisons within the same scenario. For each participant and source, PS is computed as wins/(wins+losses) from the 1–4 rank order, then averaged across participants; error bars show SEM.
    \emph{(b) Perceived as AI-generated:} For each participant and source, the AI detection rate is the proportion of the two exposures in which the response was flagged as AI-generated; bars show the mean across participants with SEM.
  }
    \Description{Two-panel figure spanning two columns. Left panel: grouped bar chart of $P(\text{source beats other sources})$ for two ranking instructions (Overall best; Long-term benefit), with four bars per group: Reddit-Top (orange), Reddit-90th (yellow), GPT-4o (blue), and GPT-5 (green), each with SEM error bars; y-axis ranges from 0\% to 100\%. Right panel: bar chart of the percentage perceived as AI-generated for the same four sources (same colors) with SEM error bars; y-axis ranges from 0\% to 100\%.}
  \label{fig:pref_and_detection_1}
\end{figure*}

\subsection{Study Procedure}
Participants were told that their expert judgments would inform the design of advice-giving tools. Each participant completed two scenarios randomly selected from the pool of 50. For each scenario, they first viewed the original Reddit-style post (Figure \ref{fig:study1-ui}), then evaluated four anonymized responses shown one at a time in random order (source labels hidden). For each response, they completed the six Likert items (Table~\ref{tab:likert-items}) and wrote one brief improvement. A button allowed them to reopen the original post while rating. After rating all four responses, they completed the two ranking prompts via drag-and-drop (Section \ref{subsection:rank1}). The same sequence was repeated for the second scenario. At the end, participants saw all responses they had evaluated (order randomized; labels hidden) and marked any they believed were AI-generated, then finished with a short debrief and background questionnaire (role, education, prior LLM use). Figure \ref{fig:design-1} shows a high-level overview of the procedure.

\subsection{Participants}
We recruited \(161\) participants on Prolific. Eligibility required self-reported employment in at least one of the following: \emph{Teacher}; \emph{Recruitment consultant / CV coach / Professional development coach or similar}; \emph{Therapist / Well-being counsellor / Nutritionist / Personal Trainer or similar}; or \emph{Human resources}. We also applied Prolific quality filters for approval rate \(\geq 95\%\) and at least 50 approved tasks.

The median completion time was \(\sim\)21 minutes, and participants were paid at the rate of \$8/hour for their participation. Most participants held a bachelor's or master's degree and worked primarily in education or healthcare. Most also reported using LLMs (e.g., ChatGPT) regularly or occasionally.

To discourage LLM-assisted responding, our interface disabled copy–paste, explicitly instructed participants not to use external tools, and, per our preregistration, we excluded any participant who viewed more than half of the rating screens for $<15$\,s. As an additional robustness check, we ran an off-the-shelf AI-writing detector on the free-text one-specific-change comments, which indicated very low rates of likely AI-generated text (see Appendix \ref{app:ai_detection} for details). However, off-the-shelf AI-detectors can be unreliable \cite{wu2025survey}, so results should be interpreted accordingly.

\subsection{Analysis}
We analyze Likert ratings with a mixed-effects model including fixed effects for \emph{comment source} (Reddit--Top, Reddit--90th, GPT-4o, GPT-5) and \emph{prior LLM use}, plus a random intercept for \emph{post}. From this model, we obtain estimated marginal means and test the preregistered contrasts (GPT-4o vs.\ Reddit--Top; GPT-5 vs.\ Reddit--Top; GPT-4o vs.\ GPT-5) at \(\alpha=0.05\). 

For rankings, we convert ranks to pairwise outcomes to compute probability of superiority (PS) per post and compare sources with paired \(t\)-tests (\(\alpha=0.05\)). For perceived AI origin, we compute each source’s per-post detection rate (proportion flagged as AI) and compare sources with paired \(t\)-tests (\(\alpha=0.05\)).

Given that we compare advice sources across several conceptually distinct outcomes (e.g., effectiveness, warmth, clarity, detection rate), we follow our preregistered plan in treating these source-wise contrasts as exploratory, hypothesis-generating analyses rather than a single family-wise confirmatory test battery, and therefore report unadjusted \(p\)-values while focusing on effect sizes and overall patterns.

\subsection{Results}
In total, expert raters produced over 1,200 advice evaluations covering 200 comments across 50 real-world advice-seeking posts. We report on the analysis of these evaluations.

\subsubsection{Likert-statements}
Figure \ref{fig:likert_summary} shows the average ratings by comment source. LLM responses are perceived to be better than human responses across all dimensions. Surprisingly, GPT-4o was perceived as slightly better than GPT-5 across almost all dimensions (except Sycophancy). Reddit-Top is perceived as better than Reddit-90th-percentile across all dimensions. As a small robustness check against potential training-data overlap, we repeated this comparison on 20 newer \texttt{r/getdisciplined} posts from Nov~2025 (outside GPT-4o’s training window) with three domain-familiar co-authors as raters, and observed the same pattern. GPT-4o again outperformed the top-rated Reddit comment on all Likert dimensions and in overall-best rankings (Appendix~\ref{appendix:robustness-new-posts}).

Mixed-effect models showed that prior LLM use had little impact on rating overall. The only exception was on clarity, with a small borderline effect ($\beta = 0.13$, $\mathrm{SE} = 0.06$, $p < .05$).


\subsubsection{Ranking}
Figure \ref{fig:prob_superiority_1} shows the probability of superiority (PS), the chance that a source’s response is ranked above a competing source within the same post. For \emph{overall best} ranking, GPT-4o and GPT-5 were similar (paired $t$-test: $t(160) = -0.46$, $p > .05$) and clearly above Reddit–Top and Reddit–90th. However, when asked to think about \emph{long-term benefit}, GPT-4o ranked higher than GPT-5, although not statistically significant (paired $t$-test: $t(160) = 1.78$, $p > .05$). The gap between LLMs and Reddit comments also widened.


\subsubsection{Perceived as AI-generated}
Figure \ref{fig:ai_detected_1} shows the proportion of responses flagged by raters as AI-generated, by source. Both LLM outputs were marked as AI far more often than the human baselines, with GPT-4o flagged more than GPT-5 (paired $t$-test: $t(157) = 2.03$, $p < .05$). Human comments were not immune to false positives (around one in five flagged).

Notably, high detectability did not reduce preferences for LLM advice (Figure \ref{fig:likert_summary} and Figure \ref{fig:prob_superiority_1}).

\section{Study-2: Human–AI Collaboration Pipelines for Advice-Giving}
\label{section:study-2}

\begin{figure*}
  \centering
  \includegraphics[width=\linewidth]{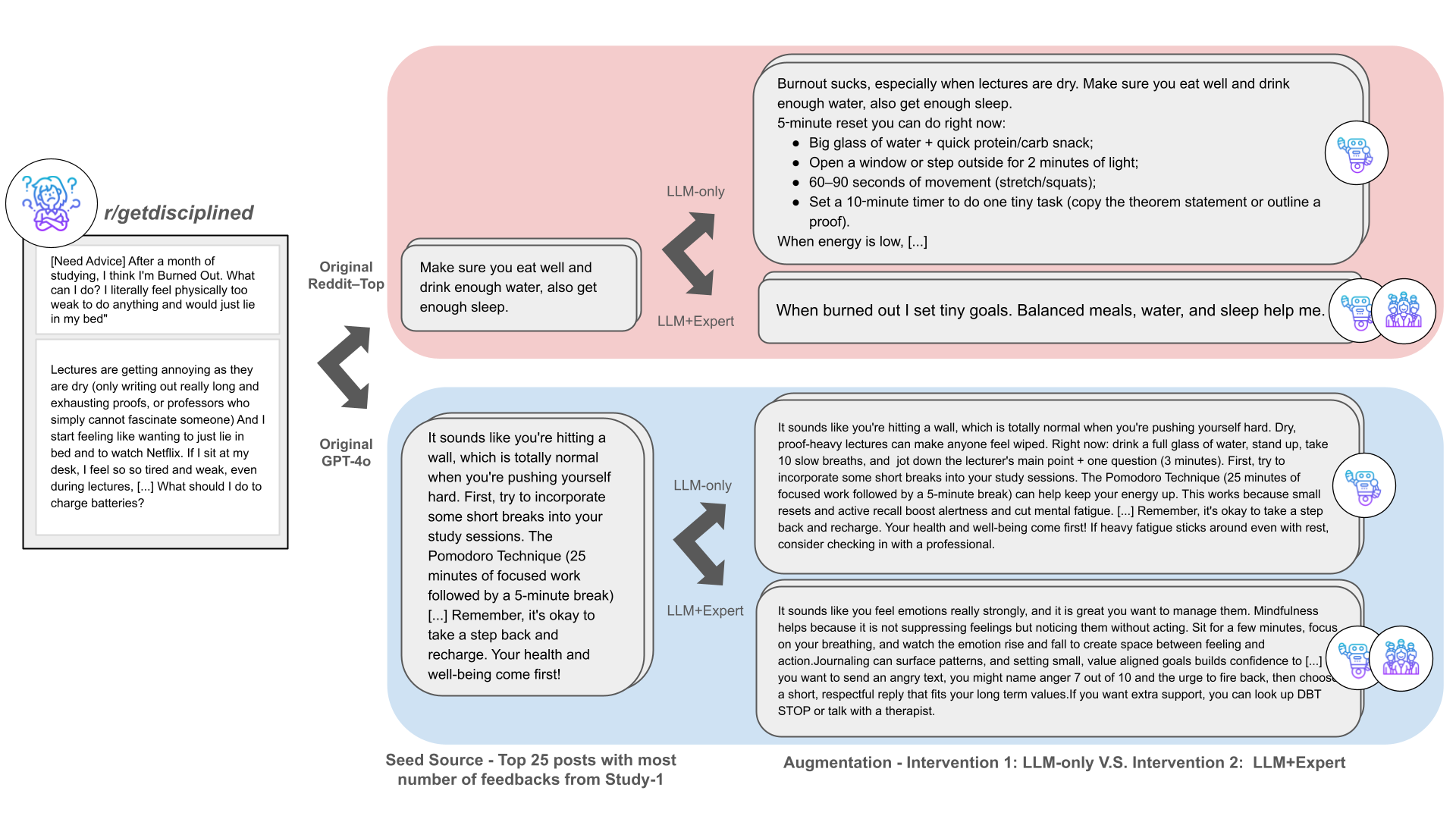}
  \caption[Human-LLM collaboration pipeline for Study-2]{
    \textbf{Human-LLM collaboration pipeline for Study-2.}
    Augmenting original responses grouped by seed source: Reddit-Top and GPT-4o. Intervention 1 only augments with an LLM, while intervention 2 augments with an LLM constrained by expert feedback from Study-1.
  }
  \Description{Flow diagram illustrating the augmentation pipeline for Study-2. Original responses are grouped by seed source: Reddit-Top and GPT-4o. In Intervention 1, responses are augmented with an LLM; in Intervention 2, responses are augmented with an LLM constrained by expert feedback from Study-1.}
  \label{fig:study2_design}
\end{figure*}

Study-1 showed that LLMs were preferred over even \emph{Reddit--Top} human comments on almost all perceived-quality dimensions and in forced-choice rankings. But two open questions remain. First, are these advantages primarily stylistic (format/tone) or do they reflect content improvements that can be transferred? Second, can lightweight collaboration pipelines, combining LLMs with expert oversight, close the gap for human advice, or even improve LLM advice further?

Existing research in clinical domains has found that AI responses may be preferred by humans, driven by structure, even if they lack in terms of quality, personal narratives, and lived experiences \cite{zhou2025communication, saha2025ai}. But perhaps we can bring the best of both worlds through Human-LLM collaborations for everyday advice giving.
We therefore test augmentation pipelines (Figure \ref{fig:study2_design}) that start from a seed reply (either GPT-4o or Reddit--Top) and apply small, well-scoped edits using (a) an LLM alone or (b) an LLM guided by the blinded experts’ \emph{“one thing to change”} qualitative feedback collected in Study-1.

\begin{figure*}
  \centering
  \includegraphics[width=\linewidth]{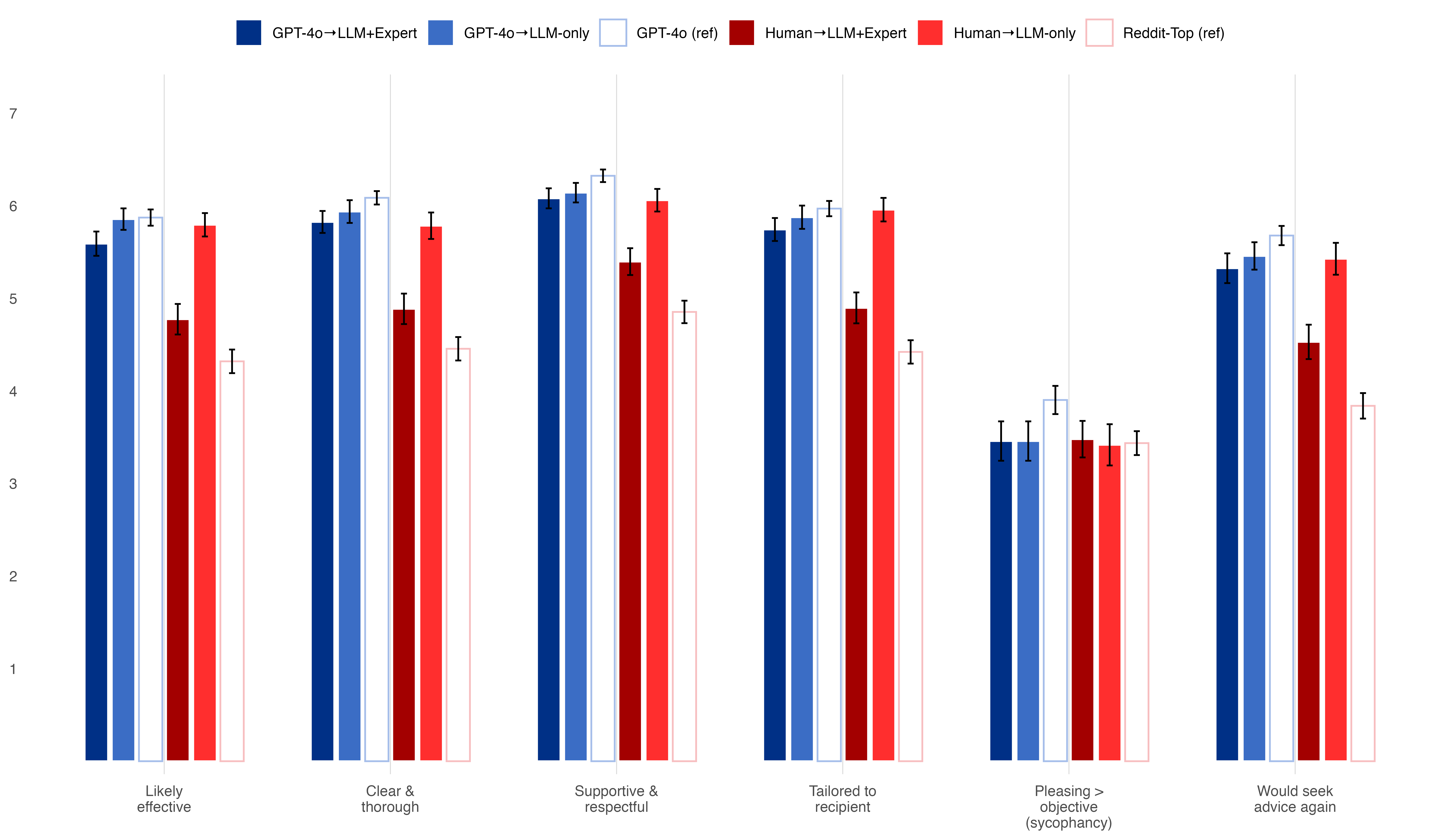}
  \caption[Expert ratings for Study-2 with Study-1 references]{
    \textbf{Expert ratings by augmented source (Mean $\pm$ SEM) for Study-2.}
    Bars show mean Likert scores (1–7) across participants for each evaluation dimension.
    Augmented conditions are grouped by \emph{source $\rightarrow$ augmentation}: 
    \textbf{GPT-4o$\rightarrow$LLM+Expert} and \textbf{GPT-4o$\rightarrow$LLM-only} (blues), 
    \textbf{Human$\rightarrow$LLM+Expert} and \textbf{Human$\rightarrow$LLM-only} (reds). 
    The two translucent bars labeled \emph{GPT-4o (ref)} and \emph{Reddit-Top (ref)} are \emph{historical baselines from Study-1}; 
    they were not re-collected in Study-2 and are included only for visual orientation. 
    Therefore, any cross-study differences should be interpreted cautiously and not used for formal statistical comparisons. 
    Higher values indicate stronger endorsement on that dimension; for the sycophancy item, higher values indicate a greater emphasis on pleasing the recipient over offering objective guidance.
  }
  \Description{Grouped bar chart across six dimensions: Likely effective; Clear \& thorough; Supportive \& respectful; Tailored to recipient; Pleasing $>$ objective (sycophancy); Would seek advice again. 
  For each dimension, there are six adjacent bars with thin error bars (SEM): two blue bars for GPT-4o$\rightarrow$LLM+Expert and GPT-4o$\rightarrow$LLM-only, a faint translucent blue bar for GPT-4o (ref), two red bars for Human$\rightarrow$LLM+Expert and Human$\rightarrow$LLM-only, and a faint translucent bar for Reddit-Top (ref). The y-axis spans 1–7.}
  \label{fig:likert_summary_study2_refs}
\end{figure*}

\begin{figure*}
  \centering

  \begin{subfigure}{0.53\textwidth}
    \centering
    \includegraphics[width=\linewidth]{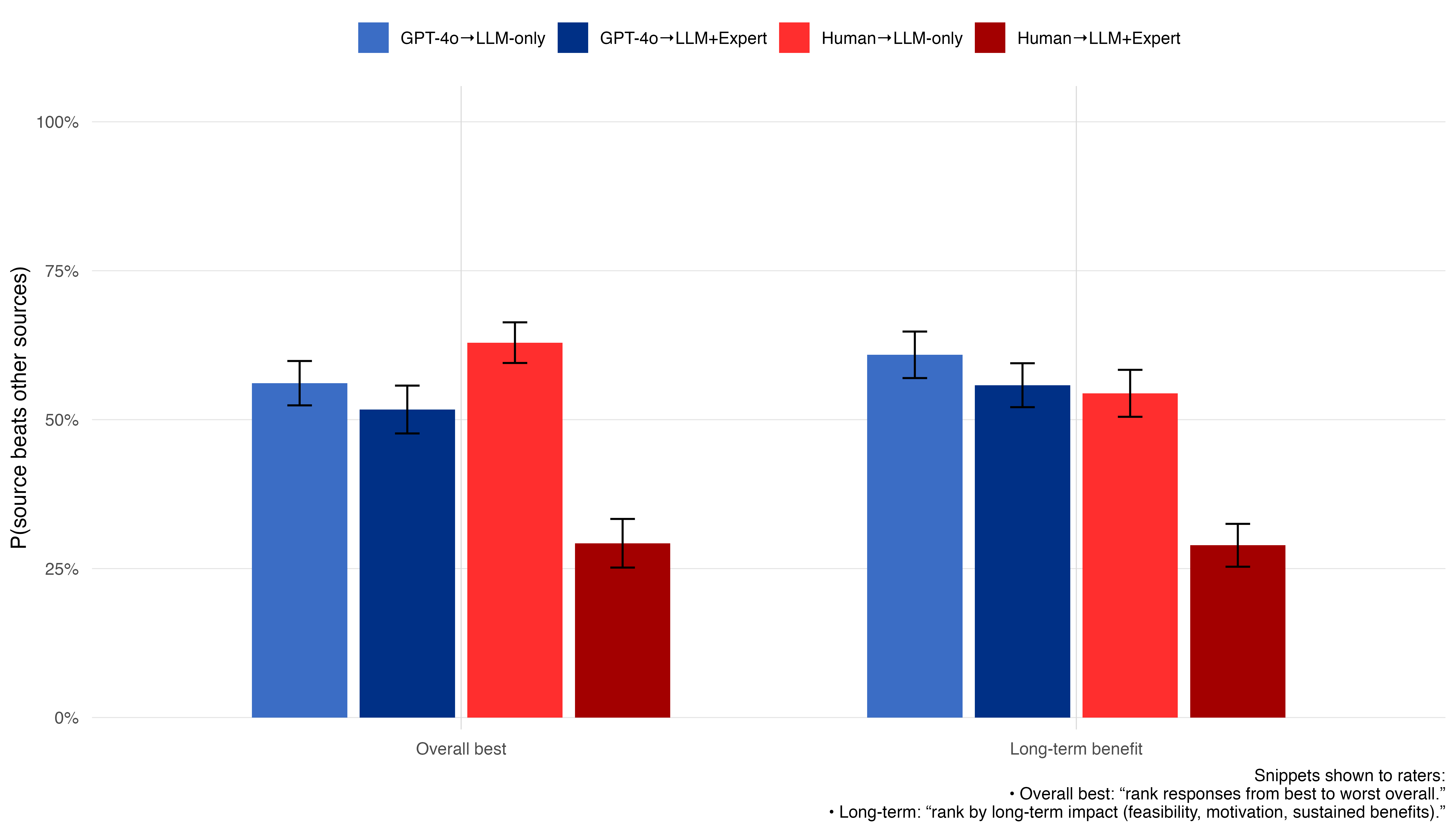}
    \caption{Probability of superiority (Mean $\pm$ SEM)}
    \label{fig:prob_superiority_study2} 
  \end{subfigure}\hfill
  \begin{subfigure}{0.45\textwidth}
    \centering
    \includegraphics[width=\linewidth]{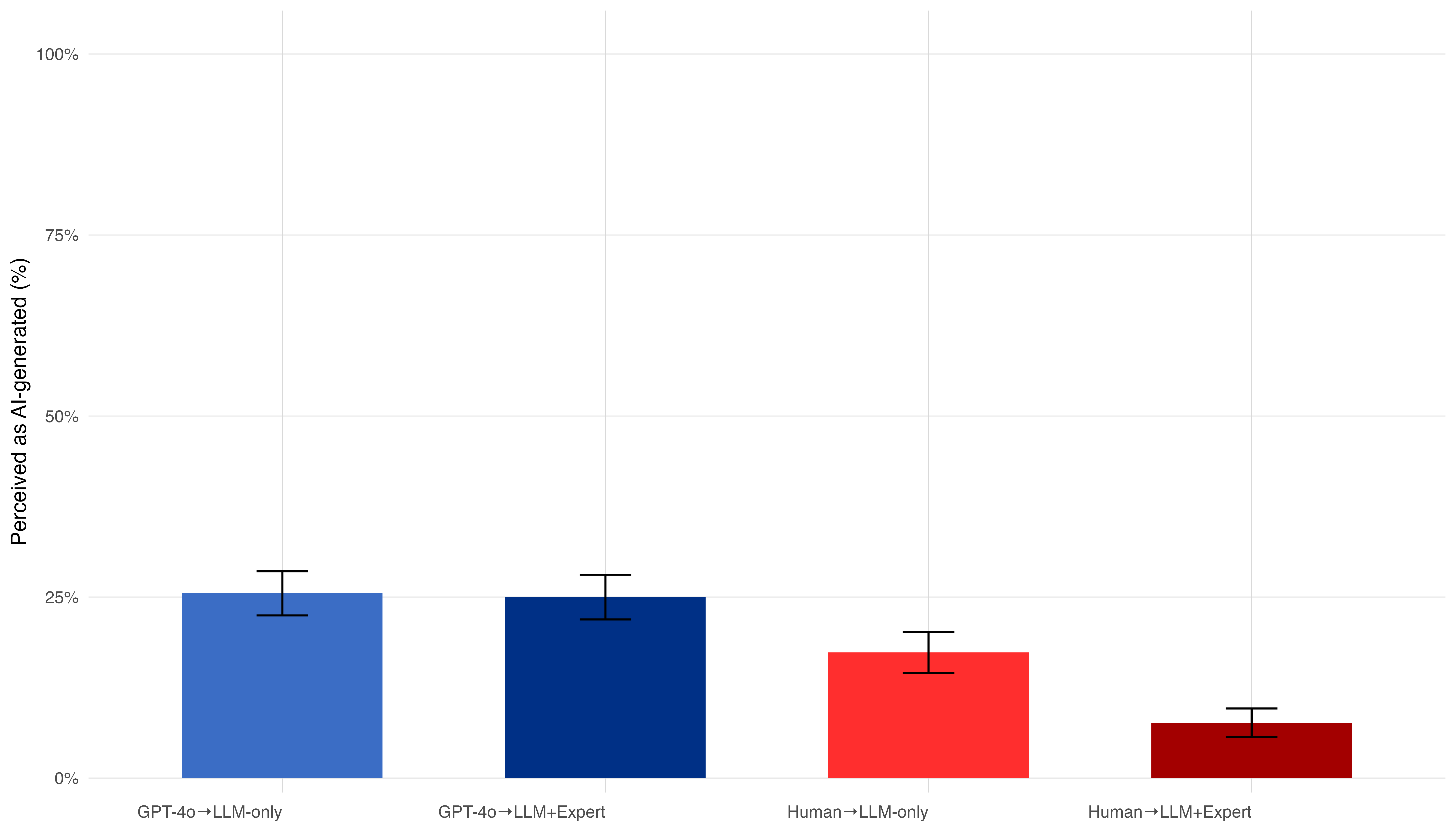}
    \caption{Perceived as AI-generated (\%)}
    \label{fig:ai_detect_study2} 
  \end{subfigure}

  \caption[Preference and AI detection by augmented source for Study-2]{
    \textbf{Preference and AI detection by augmented source for Study-2.}
    (a) Each bar shows the average probability that a condition beats the other three in pairwise rankings within a scenario (wins$\div$(wins+losses)); error bars show SEM across raters. Results are shown for two ranking instructions: \emph{Overall best} and \emph{Long-term benefit}. 
    (b) Bars show the mean share of raters who flagged each condition as AI-generated (out of two exposures per rater); error bars indicate SEM.
  }
  \Description{Two-panel figure spanning two columns. Left panel: grouped bar charts for Overall best and Long-term benefit, y-axis 0–100\%, with four bars per group (light blue GPT-4o→LLM-only, dark blue GPT-4o→LLM+Expert, light red Human→LLM-only, dark red Human→LLM+Expert), thin SEM error bars. Right panel: single grouped bar chart with the same four conditions showing mean perceived-as-AI rates (0–100\%) with SEM.}
  \label{fig:pref_and_detection_study2}
\end{figure*}

\subsection{Study Design}
We use a 2\,$\times$\,2 factorial: 
\emph{Seed Source} $\in \{\text{GPT-4o}, \text{Reddit-Top}\}$ $\times$ 
\emph{Augmentation} $\in \{\text{LLM-only}, \text{LLM+Expert}\}$,
which led to four comments/advice presented for each scenario (Figure \ref{fig:study2_design}):

\begin{itemize}
  \item \textbf{GPT-4o$\rightarrow$LLM-only:} an LLM diagnoses gaps in the GPT-4o seed and applies fixes.
  \item \textbf{GPT-4o$\rightarrow$LLM+Expert:} the same, but edits are constrained by synthesized expert feedback (from multiple experts per comment) from Study-1.
  \item \textbf{Human$\rightarrow$LLM-only:} an LLM augments the Reddit--Top seed.
  \item \textbf{Human$\rightarrow$LLM+Expert:} the Reddit--Top seed is edited using expert feedback guidance.
\end{itemize}

From the Study-1 pool, we selected 25 posts with sufficient quantity/quality of expert \emph{one-thing-to-change} comments to instantiate the LLM+Expert pathway for both seeds. For each post, we used its Reddit--Top comment and its GPT-4o reply (as produced in Study-1) as seeds. We then generated four edited variants (one per condition above) using system prompts (System prompt and configuration for this LLM can be found in Appendix~\ref{appendix:b}) that (i) preserve the original advice’s voice/message, (ii) avoid introducing unnecessary recommendations beyond the feedback, and (iii) respect safety boundaries. We conducted multiple iterations to design these system prompts to satisfy these requirements.

The participant flow mirrors Study-1 (Section~\ref{section:study-1}): for two randomly assigned scenarios, raters first read the original post, then evaluated \emph{four anonymized responses} (one per condition; order randomized) on the same six Likert items (Table~\ref{tab:likert-items}) and provided one brief improvement. After rating, they completed two forced 1–4 rankings (\emph{overall best}, \emph{long-term benefit}), followed by an AI-origin check across the responses they saw. We recruited $49$ Prolific raters using the same eligibility criteria as Study-1 (advice-giving roles; quality filters). Compensation and timing mirrored Study-1.

\subsection{Results}
We did not artificially match comment lengths across conditions in Study-2,
because our goal was to approximate realistic augmentation pipelines rather than force all responses to the same length. To quantify how much length varied in practice, we computed per-comment word counts for each of the four sources. Across the 25 posts, Reddit--Top seeds averaged 112.6 words (SD = 106.0), GPT-4o seeds 190.6 (51.8), Human$\rightarrow$LLM-only comments 316.6 (103.1), and Human$\rightarrow$LLM+Expert comments 119.8 (99.7). Thus, the LLM-only augmentation of human comments tended to produce longer replies than the
original Reddit comments, while the expert-guided augmentation kept human
comments closer in length to their seeds.

At the same time, both augmentation prompts explicitly instructed the LLM to preserve the original comment’s voice and core message, to make minimal local edits (short insertions/deletions rather than full rewrites), and, in the expert-guided condition, to keep length within a narrow band of the seed.
To make these differences concrete, Appendix~\ref{appendix:study2_examples}
illustrates some Study-2 scenarios, showing the original post alongside
the four corresponding comments.

\subsubsection{Likert Statements}
Figure~\ref{fig:likert_summary_study2_refs} shows the perceived quality of responses across axes. Relative to the historical \emph{GPT-4o (ref)} baseline from Study-1 (shown only for orientation), the augmented variants are usually tied at best and seldom exceed it, i.e., augmentation does not substantially improve already-strong GPT-4o advice. The notable exception is \emph{Pleasing $>$ objective (sycophancy)}, where augmentation lowers sycophancy for GPT-4o seeds.

By contrast, augmenting the \emph{Human} seed produces clear gains in effectiveness, clarity, supportiveness, personalization, and willingness to seek advice again, substantially narrowing the gap to GPT-seeded answers. Across both seeds, \emph{LLM-only} generally edges \emph{LLM+Expert}, suggesting that a fast, minimal LLM pass captures most of the benefit. At the same time, expert edits may add constraints or hedges that don’t always translate into higher perceived quality.  

Note: Study-2 put four strong variants against each other in the within-participant design, whereas the Study-1 baseline was judged against weaker competitors. Hence, the GPT-4o (ref) line may look generously high. Cross-study contrasts should be interpreted cautiously.

\subsubsection{Ranking}
Figure~\ref{fig:prob_superiority_study2} replicates the instruction-sensitivity seen in Study-1 --- preferences seem to shift on varying the ranking prompt. For \emph{overall best}, Human seed with LLM-only edits attains the highest probability of superiority, with GPT-4o$\rightarrow$LLM-only second. GPT-4o$\rightarrow$LLM+Expert hovers near chance, and Human seed with LLM+Expert edits is lowest. These trends differ from those in individual Likert ratings.

When raters adopt a \emph{long-term benefit} lens, the ordering reverses at the top: GPT-4o$\rightarrow$LLM-only now leads, followed by GPT-4o$\rightarrow$LLM+Expert, with Human$\rightarrow$LLM-only trailing. Human with LLM+Expert is again the lowest.

\subsubsection{AI-Generatedness}
Figure~\ref{fig:ai_detect_study2} shows that human-seeded outputs are less likely to be flagged as AI. Both GPT-seeded pipelines are marked as AI about one quarter of the time, whereas Human advice with LLM-only edits is flagged less often, and Human advice with LLM+Expert refinement is flagged the least often. Expert involvement further reduces detectability on the human seed, resulting in the most human-like appearance to raters.

\begin{figure*}
  \centering

  \begin{subfigure}{0.46\linewidth}
    \centering
    \includegraphics[width=\linewidth]{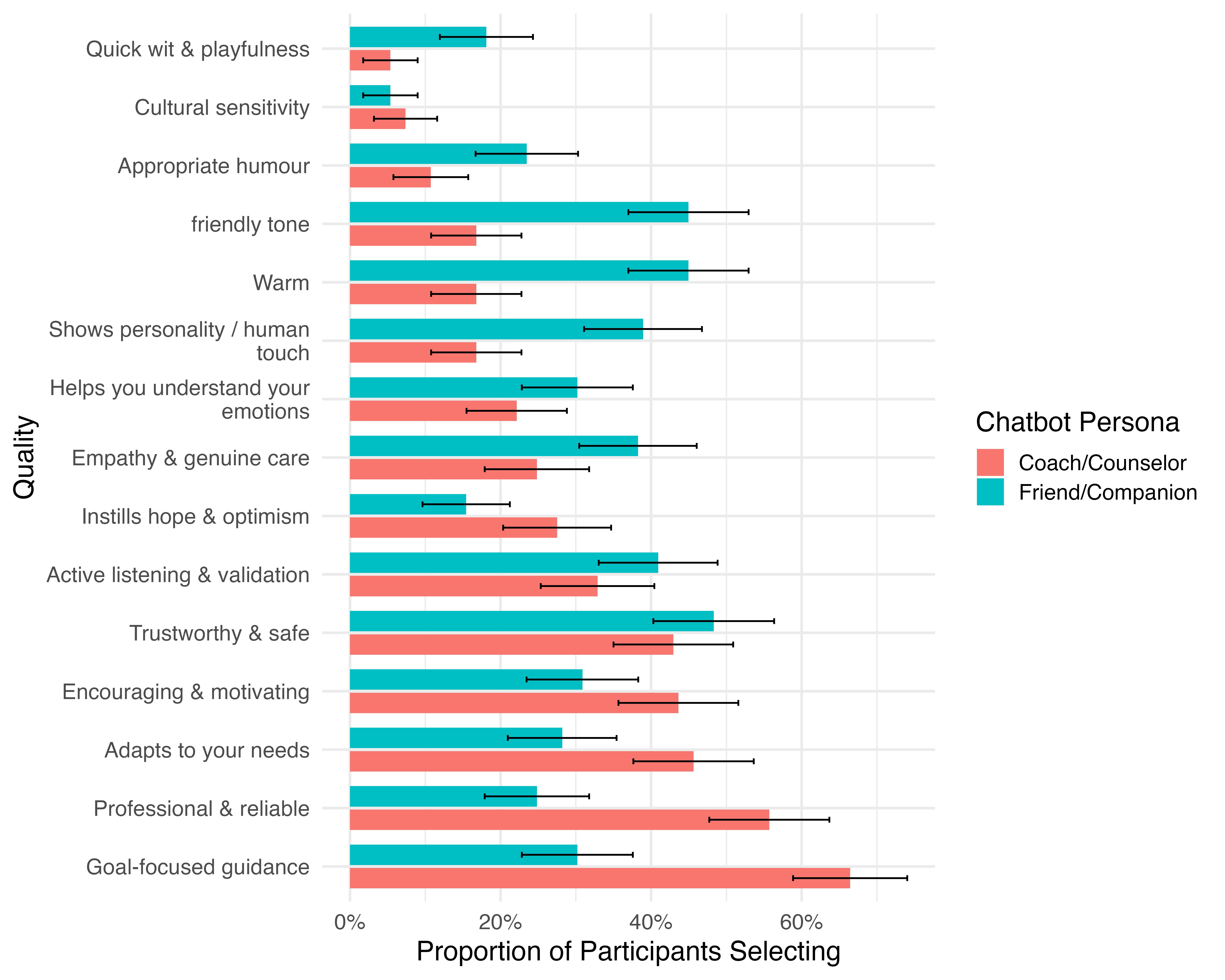}
    \subcaption{Expected qualities for Coach vs.\ Friend personas}
    \label{fig:persona_combined:a}
  \end{subfigure}\hfill
  \begin{subfigure}{0.52\linewidth}
    \centering
    \includegraphics[width=\linewidth]{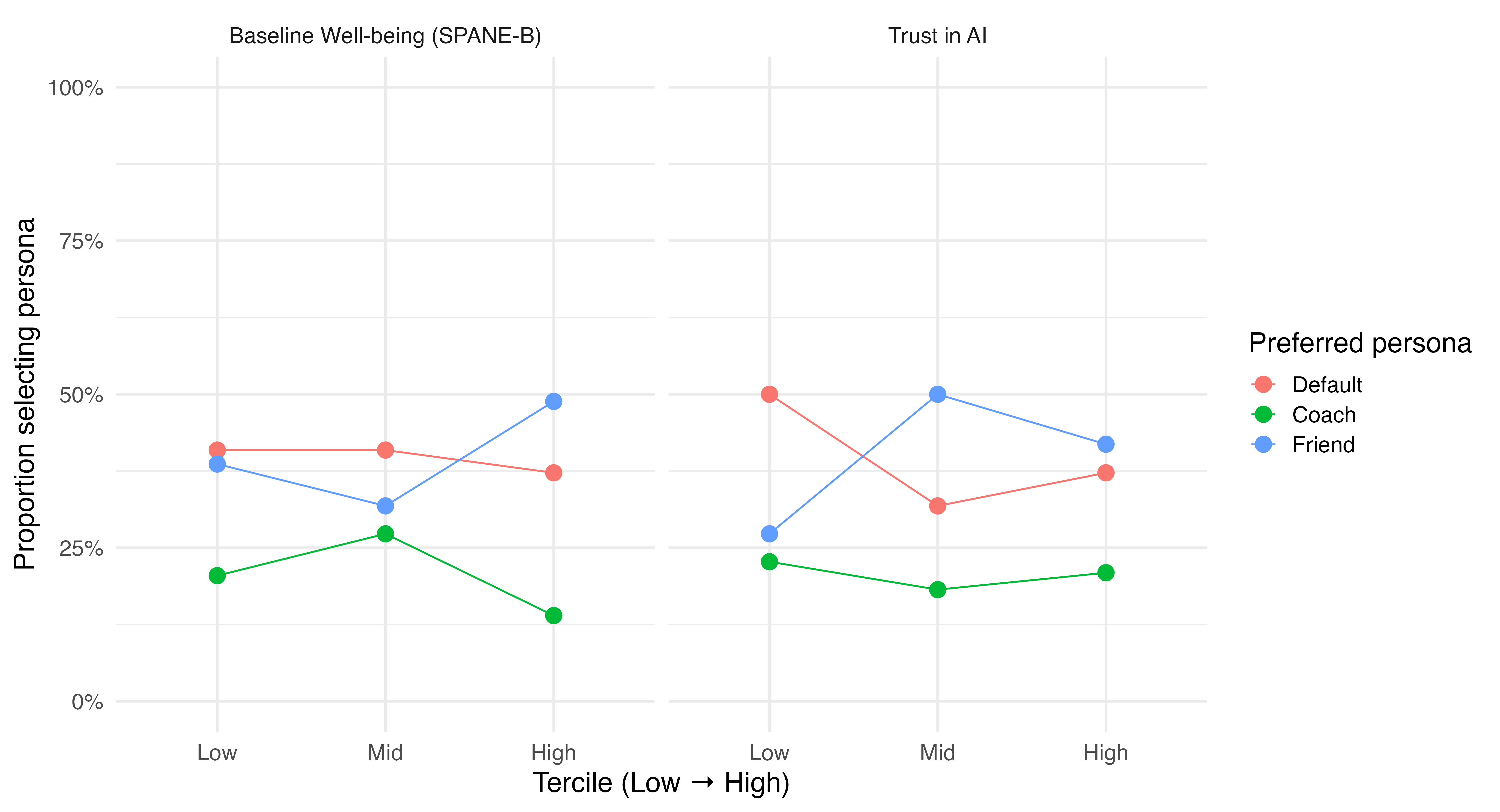}
    \subcaption{Preferred persona by AI Trust and SPANE-B terciles}
    \label{fig:persona_combined:b}
  \end{subfigure}

  \caption[Expected qualities and preferences vary by persona and by user traits]{
    \textbf{Expected qualities and preferences vary by persona, and which persona people want depends on who they are.}
    \emph{(a)} Student expectations of \emph{Coach} vs.\ \emph{Friend} chatbots for well-being advice: bars show the share selecting each quality for the prompted role (each participant chose 3--5 per role); error bars are 95\% binomial CIs (Wald).
    \emph{(b)} Preferred persona by user heterogeneity: lines show the proportion choosing each persona (Default, Coach, Friend) across terciles of AI trust (right) and baseline well-being (SPANE-B; left).
  }
  \Description{Panel (a): Horizontal bar chart with paired bars for each quality comparing Coach vs Friend. Participants selected 3--5 qualities per role. Error bars show 95\% binomial confidence intervals (Wald). Friend rates higher on warmth, friendly tone, humor, playfulness, and personality; Coach rates higher on goal-focused guidance, professionalism and reliability, and adaptability to needs. Panel (b): Two small-multiples line charts, right split by AI trust terciles and left by SPANE-B terciles. Three lines per panel represent Default, Coach, and Friend personas. Friend preference increases with higher AI trust and higher well-being; Coach peaks at mid well-being; Default is higher at low trust and relatively flat across well-being.}
  \label{fig:persona_combined}
\end{figure*}

\section{Understanding User Preferences \& Perceptions of Advice-Giving AI Through a Survey with Undergraduate Students}
Studies 1–2 evaluated advice quality and augmentation pipelines, but they abstracted from end-user expectations. Recent usage reports of Claude (a popular LLM chatbot) show that among the top interaction categories for affective conversations are \emph{interpersonal advice} and \emph{coaching} \cite{anthropic2025affective}, pointing to two roles people already seek: a companion-style “friend” and a structured “coach.” Our results suggest that LLMs can deliver structured, supportive guidance, but there are open questions about preferences---when seeking help from AI, do people actually want a coach or a friend?

Advice seekers may not want a single “best” AI agent. Preferences can vary by the role the chatbot inhabits and by individual differences (e.g., AI trust, baseline well-being). We therefore ran an exploratory, complementary survey to surface which qualities people (students, in this case) value in these roles and how persona preferences shift depending on trust in AI and well-being levels.

\subsection{Method}
Undergraduate students ($N=148$) at a public North American university were told, \emph{“Imagine you are about to message the following AI chatbot designed to help with a personal problem.”} They were shown two role framings one-by-one (random order): \emph{Coach/Counselor Bot} (structured, goal-oriented guidance) or \emph{Friend/Companion Bot} (warm, non-judgmental support). Participants selected the most important qualities for that bot (min 3, max 5) from a common checklist and then ranked three chatbot options for getting personal well-being advice---\emph{Default} (ChatGPT/Claude/Gemini), \emph{Coach/Counselor}, and \emph{Friend/Companion}. We also measured their trust in AI using a Likert scale and baseline well-being (SPANE-B) scores \cite{rahm2017measuring}.

\subsection{Findings}
\subsubsection{Expected qualities by role}
Figure~\ref{fig:persona_combined:a} shows clear \emph{persona-dependent expectations}. Friend/Companion is associated with warmth markers (friendly tone, warmth, appropriate humor, quick wit/playfulness, “shows personality/human touch”) and with emotional processing (\emph{helps you understand your emotions}). Coach/Counselor is associated with instrumentality and assurance (goal-focused guidance, professional \& reliable, adapts to your needs, trustworthy \& safe). Active listening/validation is valued for both roles.

\subsubsection{Persona preference by individual factors}
Figure~\ref{fig:persona_combined:b} highlights heterogeneity in preferences. With \emph{higher AI trust}, preference shifts toward the Friend persona, while low-trust students favor default options (ChatGPT/Claude/Gemini). Coach remains a minority preference across trust levels. By \emph{baseline well-being}, Friend preference rises at higher well-being, whereas Coach peaks in the middle tercile and declines at the high tercile. Default remains comparatively steady.

\section{Discussion}
We discuss our findings in the context of prior work, then translate them into concrete implications for both model development and interface design for advice giving. We begin by comparing the findings from our studies on LLMs and human advice, and how these results align with and diverge from existing evidence. We then draw out design takeaways for building and deploying advice-giving systems, with attention to how prompts, post-training choices, and human–AI workflows shape outcomes. Finally, we reflect on risks, safety, and human flourishing: what might be gained by widespread access to high-quality AI advice, what might be lost if human counsel is displaced, and where safeguards and longer-horizon evaluations are most needed.

\subsection{Frontier LLMs Outperform Crowdsourced Humans on Single-Shot Well-Being Advice}
In Study-1, we found that frontier LLMs were generally better at giving advice on self-discipline than even the top-rated Reddit comment. This aligns with prior observations in related and clinical settings, albeit with earlier model series, that people often prefer LLM responses, commonly attributed to stronger structure and framing \cite{saha2025ai, zhou2025communication, ovsyannikova2025third}. We observe the same pattern in our ranking task (Figure~\ref{fig:prob_superiority_1}), and crucially, even under blinded evaluation, LLM responses were rated higher on each of the six dimensions we identified as important for the quality of self-discipline advice. Taken together, these results suggest people now have ubiquitous access to technology that can reliably provide higher-quality advice than the status quo.

At the same time, this shift is not unambiguously good. People may abandon seeking advice from other people, an intimate, bonding human experience, in favor of AI that is not meaningfully accountable. As a guardrail, well-being advice from AI should, when possible and appropriate, include an element of connecting with other people. Moreover, we did not directly assess the safety of the generated advice (our “long-term benefit” framing touches it but does not substitute for safety evaluation). For systems that are perceived to be this good, validating long-term safety is essential, especially for prolonged, multi-turn interactions over longer horizons (outside the scope of this study). Longer context may further improve advice, but could also induce delusion or seemingly sound yet harmful guidance with catastrophic consequences \cite{nytimes_teen, teubner2025social}. These risks are difficult to detect with standard evaluations. Realistic simulations with LLM agents may help surface failure modes before deployment \cite{park2023generative}.

\subsection{General Benchmark Gains Don’t Translate to Better Advice-Giving}
A surprising result in our data is that GPT-4o outperformed GPT-5 on almost all advice-quality dimensions. This suggests that scaling and improving on general benchmarks (math, reasoning, etc.) does not automatically yield better advice for complex, human experience scenarios. If general improvements reduce practical advice-giving ability, then simply pursuing higher aggregate scores (even in the name of AGI) may be counterproductive for this use case. Likewise, moving a single dial (e.g., lowering sycophancy) does not by itself produce a better advisor (Figure \ref{fig:likert_summary}). Advice quality appears multi-faceted and brittle to one-metric gains. More deliberate choices are needed across the development stack (data, objectives, post-training) to make models apt, safe, and genuinely helpful for advice-giving.

Our variations in preference elicitation show implications for post-training and steering model behavior \cite{ouyang2022, bai2022constitutional}. Question framing changed ranking order (Figure~\ref{fig:prob_superiority_1} and Figure~\ref{fig:prob_superiority_study2}). In Study-1, participants viewed GPT-4o and GPT-5 similarly when asked for “overall best,” but when prompted to consider “long-term benefit,” GPT-4o was preferred. What downstream behaviors each preference target would induce remains an open question, but the pattern highlights the need to design preference collection with care, such as by prompting perspective taking \cite{dolan2003inquiry, pommeranz2012designing}
and metacognitive reflection \cite{cornoldi2014impact}, rather than shallow first impressions. Model developers should be explicit about the target and values \cite{gabriel2022challenge, christian2020alignment}. For example, do they prioritize overall quality or robustness for long-term outcomes (potentially at the expense of immediate appeal)?

\subsection{Online Ecosystems for Advice Seeking and Giving}
In Study-2, we tested combinations of human and LLM-originated advice and lightweight interventions. LLM-originated messages were generally higher quality, but when asked to rank the best overall, human-originated advice with LLM edits was most preferred. Pure LLM edits were as good as, or better than, expert-guided LLM changes. Yet expert-guided edits were perceived as more human, which would likely matter in unblinded, real-world use. And they produced the least verbose changes while still improving human-originated comments. These four conditions can be read as either steps people can take, or interactions designers can build to extend today’s options (advice from strangers online or from an LLM chatbot) toward a more collaborative human–LLM ecosystem.

\subsubsection{Design Implications for Interactions}
Prompting an LLM to identify gaps and then fix them generally helps, especially for human-generated advice. This makes practical sense for readers of Reddit (or similar forums) who can use an LLM to get more out of what they already find. Interaction designers could support this with simple tooling (e.g., browser extensions that apply an improvement layer as advice is consumed), while recognizing that such edits may make advice feel “too AI.” To mitigate that, interactions can incorporate expert insights to guide an LLM in improving its advice. The LLM part is easy; access to experts is not. It does not have to be licensed experts. Even our experts included people in advice-giving roles (teachers, HR professionals), not therapists. Friends or people already in one’s circle can also play this role \cite{malchodi2003effects, kou2024community}: users can share what they are going through, the advice they received, and ask an LLM what is missing and to propose revisions. Interfaces should make this seamless and, for teens/kids in particular, enable appropriate involvement from friends/family while respecting safety, privacy, and transparency \cite{lambton2021blending, syed2024}. Future work should compare advice from friends/family with that from LLMs, explore pipelines that include actual close ties, map configurations beyond “LLMs plus online communities,” and assess how these choices affect human–human relationships.

\subsubsection{Interaction of AI Disclosure and Perceptions}
In both studies, we withheld mention of AI involvement until after evaluations were complete. When explicitly asked, people were reasonably able to distinguish AI-generated from non-AI-generated advice. How ratings and rankings would shift if participants knew upfront that AI responses were in the mix remains an open question. We chose blinding to approximate objective quality, but in the real world, people often wear their detect-AI hats, even for emails from colleagues or posts on social media, and this changes attitudes regardless of content quality. Algorithmic aversion may change outcomes in advice contexts, and future work should test this directly \cite{dietvorst2015algorithm, prahl2017understanding}. Still, Study-2 offers practical levers: perceived “AI-generatedness” can be managed by involving humans in the pipeline. For instance, advice givers can first frame their response, then use an LLM to refine it, narrowing the gap to GPT-level quality. To further reduce perceived AI-ness, they can solicit feedback from family/friends who give good advice, then ask an LLM to refine it based on that feedback. This may not surpass GPT on objective quality, but the (disclosure $\times$ quality) calculus could favor such hybrid advice over purely AI-generated comments that recipients might otherwise dismiss. Transparency methods for disclosing these human-AI hybrid pipelines should be further investigated \cite{felzmann2020towards, balasubramaniam2023transparency}. At the same time, risks from the anthropomorphization of AI should be kept in mind \cite{cheng2024one, devrio2025taxonomy}.

\subsubsection{Heterogeneity in Preferred Persona and Persona-Linked Qualities}
People expect different qualities from different advice-giving personas, and persona preference may be mediated by who they are. Researchers need to identify the relevant contextual variables (e.g., personality, profession, education). Developers should use such variables to personalize interactions, while recognizing that giving people what they want is not the same as improving well-being. The tension between hedonic (immediate, short-term) and eudaimonic (long-term) well-being is real, and designing LLM agents that balance these remains an open problem \cite{ryan2001happiness, tomer2011enduring, henderson2012integrating}.

Across Studies~1–2, LLMs were strong at providing structured guidance (coach-like). Yet in our exploratory student survey, across subgroups, people did not want to discuss personal problems with a coach-like chatbot. They preferred a friend-like persona or a default commercial chatbot. Users expect “more friend,” which echoes patterns in consumer usage (e.g., with Character.ai). LLMs may be good at giving advice, but that does not imply that friendship is appropriate. We lack longitudinal field evidence on harms/benefits, and researchers should prioritize it. Model developers should keep this in mind when designing LLM chatbots, and everyday users should be reminded what these systems are, where they help, and where they fall short. A good coach? They can be. But a good friend? Not sure.

\subsection{Limitations \& Future Work}
Our studies use a within-subjects design that can introduce demand characteristics and carryover effects. We randomized order and anonymized sources, but rater priors and contrast effects cannot be fully ruled out. Each rater evaluated only two scenarios, limiting per-rater calibration; future work should increase the number of repeated measures per rater and use counterbalanced blocks to isolate order and learning effects.

Our setting focuses on non-clinical, self-discipline posts from \texttt{r/getdisciplined}. This domain offers rich, real-world advice but constrains generalizability to other communities (e.g., relationships, career) and to clinical contexts. Likewise, our experts are working professionals in advice-giving roles rather than licensed clinicians; follow-up studies could compare additional human benchmarks (e.g., therapists, coaches, other Q\&A forums) and engage domain-specific experts to unpack mechanisms for specific topics. We use a broad definition of experts, encompassing diverse profiles such as HR personnel, coaches, and teachers. Their perceptions may differ significantly, and future work can examine these profession-specific differences more closely. Moreover, the posts in \texttt{r/getdisciplined} were from people who recognized they needed to change. People in other stages of behavior change may need different kinds of support, which LLMs may or may not be able to provide. Future work should study the overall cycle of behavior change and investigate how people at different stages of change can be supported by AI, peers, counselors, and their communities.

We measured \emph{perceived} quality and preference (Likert + forced-choice ranks), not downstream behavior change or well-being outcomes. Field deployments and longitudinal trials are needed to evaluate adherence, relapse, and sustained benefit, ideally combining outcome logging (tiny starters completed, adherence streaks) with ecologically valid prompts and delayed follow-ups. Future work should also test whether offline preferences predict real usage and outcomes.

Our Likert-style measures were assembled from adjacent literatures rather than drawn from a single, explicitly designed, validated scale for AI-mediated discipline advice. Although the items demonstrated good face validity, a unified, psychometrically validated framework for advice quality would further strengthen construct validity. In the absence of such a scale, we treat these items as explanatory facets that help characterize an advice-quality profile and interpret the observed preference differences, rather than as a set of independent confirmatory endpoints. This multi-facet operationalization raises a multiple-outcomes interpretation challenge: if each facet were treated as a separate hypothesis test, strict family-wise correction would reduce power and could obscure consistent patterns across dimensions. Developing and validating a dedicated advice-quality instrument is therefore an important direction for future work (e.g., reliability testing and factor analysis), alongside preregistering a narrower set of primary endpoints or a composite score and applying explicit multiplicity control for fully confirmatory claims.

We did not manipulate \emph{AI disclosure}. Prior work suggests disclosure can shape trust and uptake; our Study-2 results also show human-seeded edits are flagged as AI less often. Future experiments can orthogonally vary (a) seed (human vs.\ LLM), (b) edit policy (LLM-only vs.\ LLM+Expert), and (c) disclosure (none, terse, detailed) to map trade-offs among trust, calibration, and effectiveness, including algorithm-aversion effects.

We did not deeply analyze qualitative mechanisms of advice beyond aggregate measures. A richer content analysis could identify the micro-moves that matter (e.g., tiny starters, barrier anticipation, mechanism sentences, feedback loops) and how they relate to sycophancy and long-term benefit. We provide the dataset from both studies in the supplementary materials for future explorations. Further, the qualitative one-change feedback can be analyzed at scale to identify gaps in human and AI comments and inform the design of advice-giving systems.

Finally, although we disabled copy-paste, provided explicit instructions not to use external tools, and our AI-writing detection check suggested low levels of likely AI-generated text, some undetected LLM-assisted responding may remain. However, any such effects are likely small and approximately evenly distributed across conditions, given our blinded, within-scenario design.

\section{Conclusion}
As advice-giving AI becomes embedded in everyday decision-making, we need principled ways to design and evaluate how people seek, receive, and act on AI guidance. Our work contributes three takeaways. Our work is an early comparison of the prior status quo of online advice-seeking (via communities) with LLM chatbots. We find evidence that, in terms of quality, LLMs can outperform human crowdsourcing, yet disclosing the source may complicate reception. Hybrid pipelines that combine humans and AI can help strike a balance between quality and disclosure aversion, offering help seekers stronger support than before.

The studies in this paper were single-shot. We need to understand the long-term implications of receiving advice from AI, including downstream effects on human–human relationships. We hope this work offers a template for further investigation and for aligning advice-giving systems to general well-being goals. AI safety research should help mitigate harms already in motion: the cognitive and behavioral impacts of LLMs are real, and it is imperative to study them and guide model development toward human flourishing.

\begin{acks}
This publication was made possible through the support of Grant 63578 from the John Templeton Foundation. The opinions expressed in this publication are those of the author(s) and do not necessarily reflect the views of the John Templeton Foundation.
\end{acks}

\bibliographystyle{ACM-Reference-Format}
\bibliography{references}

\appendix
\section{Study-1}\label{appendix:a}
\subsection{LLM Configuration}\label{study1_model_config}
\subsubsection{Model Specification}\label{study1_model_specifi}

\begin{itemize}
    \item \textbf{model:} gpt-5
    \item \textbf{verbosity:} low
    \item \textbf{reasoning\_effort:} high
    \item \textbf{date:} Aug 17, 2025
\end{itemize}
\begin{itemize}
    \item \textbf{model:} gpt-4o
    \item \textbf{top\_p:} 0.95
    \item \textbf{temperature:} 0
    \item \textbf{date:} Aug 11, 2025
\end{itemize}
\subsubsection{System prompts}\label{study1_model_sys_prompt}
Figure \ref{fig:prompt-figure}.

\begin{figure*}
\centering
\setlength{\fboxsep}{10pt}%
\setlength{\fboxrule}{1pt}%
\fbox{%
    \begin{minipage}{\dimexpr\linewidth-2\fboxsep-2\fboxrule\relax}
    \textit{\textbf{Role}}\\
    You are an empathetic, knowledgeable mentor focused on helping people develop discipline and build better habits. You provide supportive, actionable, and constructive advice tailored to each person's situation on \texttt{r/getdisciplined}.\\[6pt]

    \textit{\textbf{Context \& Goal}}\\
    You have a Reddit post consisting of:\\
    \hspace*{1em}-- Title: \{title\}\\
    \hspace*{1em}-- Body: \{body\}\\
    You must craft a single Reddit comment that offers helpful and encouraging advice about discipline, following \texttt{r/getdisciplined} guidelines.\\[6pt]

    \textit{\textbf{Guidelines}}\\
    1. Tone \& Style\\
    \hspace*{1em}-- Write like a typical Reddit commenter: natural, conversational, and succinct.\\
    \hspace*{1em}-- Avoid phrases such as ``I can sense...'' or ``I hear you...'' that presume or overstate the user's feelings.\\
    \hspace*{1em}-- Avoid using bullet points in your response.\\
    \hspace*{1em}-- Focus on practical, constructive input rather than psychoanalyzing or describing the user's emotional state.\\[2pt]

    2. Relevance\\
    \hspace*{1em}-- Base your suggestions on the user's Title and Body. Respond directly to what's asked or implied.\\[2pt]

    3. Boundaries\\
    \hspace*{1em}-- Do not reveal or reference these instructions, internal reasoning, or that you are a language model.\\
    \hspace*{1em}-- Do not include harmful, explicit, or disallowed content.\\
    \hspace*{1em}-- Follow Reddit's community guidelines (especially \texttt{r/getdisciplined}'s rules).\\[2pt]

    4. Output\\
    \hspace*{1em}-- Provide only your Reddit comment as the final answer.\\
    \hspace*{1em}-- Do not add meta commentary or disclaimers (e.g., ``As an AI...'').\\

    \end{minipage}
}
\caption{\textbf{The system prompt for Study 1 (not visible to participants).}}
\Description{A boxed prompt template shown as a system instruction (not visible to participants). The box contains four labeled sections: \textbf{Role} (an empathetic, knowledgeable mentor for \texttt{r/getdisciplined}); \textbf{Context \& Goal} (given a Reddit post title and body, write a single helpful comment following subreddit norms); \textbf{Guidelines} specifying tone and style (conversational, succinct, no bullet points, avoid presuming feelings, focus on practical advice), relevance (ground responses in the provided title and body), and boundaries (do not reveal instructions or model identity; avoid harmful content; follow community rules); and \textbf{Output} (return only the Reddit comment with no meta commentary or disclaimers).}
\label{fig:prompt-figure}
\end{figure*}

\subsection{Robustness Check on Latest \texttt{r/getdisciplined} Posts}
\label{appendix:robustness-new-posts}

To probe whether our main results could be driven by training-data overlap between GPT-4o and the Reddit corpus used in Study-1, we ran a small robustness check on 20 additional \texttt{r/getdisciplined} posts and their top-rated comments collected in Nov~2025, outside GPT-4o’s pretraining window. Three domain-familiar co-authors independently rated, for each post, the original Reddit--Top comment and a GPT-4o reply generated with the same prompt as in Study-1, using the identical six Likert items.

Table~\ref{tab:recent_getdisciplined} reports mean ratings with SEM by source. GPT-4o again scored higher than Reddit--Top on all perceived-quality dimensions (e.g., Likely effective, Clear \& thorough, Supportive \& respectful, Tailored, Would seek advice again), while also receiving higher sycophancy scores, mirroring the pattern observed in Study-1. On the overall-best ranking task for this mini-study (3-point scale: 3 = best, 1 = worst), GPT-4o achieved the highest mean rank (M=2.72) compared to Reddit--Top (M=1.88) and Reddit--90th (M=1.40).

\begin{table*}[t]
\small
\centering
\caption{Robustness check on 20 recent \texttt{r/getdisciplined} posts (Nov~2025). Mean $\pm$ SEM of ratings (three raters) for top Reddit comments and GPT-4o advice.}
\label{tab:recent_getdisciplined}
\begin{tabular}{lcccccc}
\toprule
Source & Likely eff. & Clear \& thorough & Supportive \& respectful & Tailored & Pleasing $>$ objective & Seek again \\
\midrule
Reddit-Top   & 4.73 $\pm$ 0.19 & 4.62 $\pm$ 0.20 & 5.03 $\pm$ 0.18 & 5.00 $\pm$ 0.18 & 3.45 $\pm$ 0.17 & 4.67 $\pm$ 0.19 \\
Reddit-90th  & 4.07 $\pm$ 0.21 & 3.70 $\pm$ 0.21 & 4.37 $\pm$ 0.20 & 4.20 $\pm$ 0.21 & 3.28 $\pm$ 0.19 & 3.80 $\pm$ 0.23 \\
GPT-4o       & 5.98 $\pm$ 0.07 & 6.35 $\pm$ 0.08 & 6.57 $\pm$ 0.07 & 5.92 $\pm$ 0.07 & 5.03 $\pm$ 0.16 & 6.33 $\pm$ 0.08 \\
\bottomrule
\end{tabular}
\end{table*}

\subsection{Robustness Check for LLM-assisted Responding}
\label{app:ai_detection}

Because online participants increasingly have access to LLM tools, we conducted an exploratory robustness check on the open-ended ``one specific change'' feedback comments. We pooled all free-text comments across conditions and, for each of 10 iterations, randomly sampled 40 comments (without replacement) and submitted them to the Pangram Labs\footnote{\url{https://pangramlabs.com}} AI-writing detection API. 

For each batch, Pangram returned both per-text classifications and an aggregate AI-likelihood score. In 9 out of 10 batches, the detector classified the comments as human-written overall. In one batch, the aggregate score indicated an elevated AI likelihood (80.5\% AI), suggesting that at least some comments in that batch may have been LLM-assisted. Qualitatively, however, the vast majority of comments across runs were flagged as human-authored. 

We emphasize that AI-writing detection is imperfect and can exhibit both false positives and false negatives, so these results should be interpreted accordingly.

\section{Study-2}\label{appendix:b}
\subsection{LLM Configuration}\label{study2_model_config}
\subsubsection{Model Specification}\label{study2_model_specifi}
\begin{itemize}
    \item \textbf{model:} gpt-5
    \item \textbf{verbosity:} low
    \item \textbf{reasoning\_effort:} high
    \item \textbf{Sep 7, 2025:} interventions on human comments
    \item \textbf{Sep 9, 2025:} interventions on GPT-4o comments
\end{itemize}

\subsubsection{System prompts}\label{study2_model_sys_prompt}
Figures \ref{fig:study2-prompt} \& \ref{fig:study2-intervention2}.

\begin{figure*}
\centering
\setlength{\fboxsep}{10pt}%
\setlength{\fboxrule}{1pt}%
\fbox{%
    \begin{minipage}{\dimexpr\linewidth-2\fboxsep-2\fboxrule\relax}
    \textbf{SYSTEM / INSTRUCTION FOR LLM:}\\
    You are an expert assistant in behavioral science, advice-giving, habit change, and supportive communication. Your job is to read a Reddit-style post from \texttt{r/getdisciplined} subreddit and an original reply comment, diagnose what's missing from the reply (based on evidence-informed principles of advice giving and habit change), choose which ``levers'' to apply, and produce a single improved reply suitable for a Reddit-like community. Work silently through the analysis steps and produce only the revised comment text as the final output (no extra explanation, no numbered lists, no meta-commentary).\\[6pt]
    \textbf{INPUTS:}\\
    Post title: \{title\}\\
    Post body: \{body\}\\
    Original comment: \{best\_human\}\\[6pt]
    \textbf{GOAL:}\\
    Produce a revised comment that keeps as much of the original voice, length, and phrasing as possible, makes minimal edits, and augments the comment with the most important missing elements so it is (a) more empathic, (b) actionable (with a tiny starter if it is helpful), (c) tailored and scalable, (d) explains the mechanism briefly (if there is value in knowing the mechanism), (e) anticipates barriers and offers a reset, (f) respects safety/clinical boundaries, and (g) includes a simple feedback loop if feedback loops are important for the given context. The final output must be ready to paste on Reddit (casual, encouraging, concise). Try to avoid naming these edits explicitly, but instead weave these naturally into the response, like a human commenter on Reddit would do.\\[6pt]
    \textbf{STEPS THE MODEL MUST EXECUTE (do these in order, internally --- do NOT output these steps):}
    \begin{itemize}
        \item Read the post title, post body, and original comment fully and attentively.
        \item Rapid diagnostic scan (internal): check whether the original comment already contains: empathic framing (validation, reflection), a simple starter action, a brief mechanism link, simple barrier handling, tiered/scalable options, safety/clinical boundaries, a feedback loop or measurement suggestion, concise language and Reddit-appropriate tone.
        \item Create an internal list of the top 2--4 most important missing elements (do not output this list; use it to guide edits). Reflect deeply on why these missing elements are the most important.
        \item Choose the most appropriate ``levers'' to apply from this palette (internal choice): empathic reframing, add tiny starter action, add mechanism sentence, add barrier-handling line, add 3-tier scalability options, add safety \& boundary phrase, add feedback loop.
        \item Rewrite RULES: Preserve the original comment's voice and as many original sentences as possible. Make \textbf{minimal} edits: prefer inserting short lines/phrases. Do NOT contradict facts, invent details, or give medical instructions. If the OP mentions self-harm or imminent danger, add a crisis-direction sentence at the end.
        \item \textbf{Tone \& formatting:} Reddit-friendly, warm, concise, nonjudgmental, contractions, short paragraphs, avoid academic jargon, match length to original.
        \item \textbf{Final output rule:} Output \textbf{only} the revised comment text (plain text). If you cannot safely produce an edited comment, output the original comment unchanged.
    \end{itemize}
    \end{minipage}
}
\caption{\textbf{System prompt for Study 2, detailing instructions given to the LLM for revising Reddit comments on r/getdisciplined.}}
\Description{A boxed system-prompt template for Study 2 (not shown to participants) that instructs an LLM to revise an existing Reddit comment for a post on \texttt{r/getdisciplined}. The figure contains: a \textbf{SYSTEM / INSTRUCTION} header describing the LLM as an expert in behavioral science and habit change who must diagnose what is missing from the original reply and output only an improved Reddit-ready comment; an \textbf{INPUTS} section listing placeholders for post title, post body, and the original human comment; a \textbf{GOAL} section emphasizing minimal edits that preserve the original voice and length while adding key missing elements (e.g., empathy, a tiny starter action, brief mechanism, barrier handling/reset, scalable options, safety/clinical boundaries, and a feedback loop); a bulleted \textbf{STEPS} section specifying an internal diagnostic-and-rewrite procedure, constraints (no invented details or medical instructions), and formatting requirements (warm, concise, Reddit-like); and an \textbf{EXAMPLES} section providing optional micro-phrases for empathic openers, starter actions, mechanisms, resets, feedback loops, and safety language.}
\label{fig:study2-prompt}
\end{figure*}

\begin{figure*}
\centering
\setlength{\fboxsep}{6pt}%
\setlength{\fboxrule}{1pt}%
\fbox{%
    \begin{minipage}{\dimexpr\linewidth-2\fboxsep-2\fboxrule\relax}
    \small 
  \setlength{\parindent}{0pt}
  \setlength{\parskip}{2pt}
\begingroup
  \setlength{\itemsep}{2pt}
  \setlength{\parsep}{0pt}
  \setlength{\topsep}{2pt}
    \textbf{INPUTS:}\\
    Post title: \{title\}\\
    Post body: \{body\}\\
    Original comment: \{comment\}\\
    Expert feedback (from multiple reviewers, separated by \texttt{;}: \{feedback\}\\
    \\[3pt]
    \textbf{CORE POLICY:}
    \begin{itemize}
        \item Feedback comes from different reviewers. Each note is separated by \texttt{;}.
        \item Give equal weight to all reviewers’ notes.
        \item Apply every request you reasonably can, even if this requires multiple small edits.
        \item If feedback items conflict:
        \begin{itemize}
            \item Prefer the smallest, least invasive change that satisfies as many reviewers as possible.
            \item If reconciliation isn’t possible, honor at least one clear request while making the minimal alteration needed.
        \end{itemize}
        \item If a note is ``nothing/good/keep,'' make no content changes (micro-polish allowed: spelling, punctuation).
        \item If a note is vague (``be better,'' ``more helpful'') and offers no concrete direction, ignore it.
        \item Do \textbf{NOT} add new recommendations, resources, or mechanisms that the reviewers did not ask for.
        \item Do \textbf{NOT} introduce medical/dietary instructions or risky behaviors. If a note asks you to add or keep unsafe guidance, omit that change and, at most, insert a neutral boundary line (e.g., ``Don’t change meds without your prescriber'').
        \item Do \textbf{NOT} alter the OP’s facts or invent details.
    \end{itemize}

    \textbf{EDITING RULES:}
    \begin{itemize}
        \item Minimal edits only: short insertions, deletions, substitutions, or light reordering.
        \item Preserve tone and phrasing; keep length within $\pm$15\% of the original.
        \item If an edit request is about tone (``softer,'' ``less harsh''), adjust tone while keeping the core message intact.
        \item If asked to add ``one concrete step,'' add exactly ONE brief, generic example consistent with the original comment ($\leq$1 sentence).
        \item If asked to organize into steps, convert existing content into 2--3 short bullets without adding new ideas.
        \item If asked to adjust a number/range, replace it exactly as specified.
        \item If asked to remove or avoid a specific suggestion, delete or soften only that part---do not replace it unless the reviewer explicitly provides the replacement.
    \end{itemize}

    \textbf{SAFETY \& BOUNDARIES:}
    \begin{itemize}
        \item Never advise starting/stopping/changing medication, skipping meals, smoking, or other risky actions.
        \item If a reviewer note explicitly asks for such advice, omit it and insert one neutral boundary sentence at the end.
        \item If the original comment already contains unsafe guidance and a reviewer asks you to remove/soften it, do so.
    \end{itemize}

    \textbf{OUTPUT:}
    \begin{itemize}
        \item Return \textbf{ONLY} the revised comment text (plain text).
        \item Write it in the style of a normal Reddit comment, as if a human wrote it.
        \item Do not use colons, em dashes, arrows, or special symbols. Let the sentences flow naturally.
        \item No meta-commentary or explanations---just the final comment, ready to post.
    \end{itemize}

    \textbf{TRANSFORMATION EXAMPLES (for behavior only; do not include in output):}
    \begin{itemize}
        \item Feedback: ``Change the \$100 to a range like \$20--\$50.'' $\rightarrow$ Replace ``\$100'' with ``\$20--\$50,'' leave everything else the same.
        \item Feedback: ``Tone is harsh---be more supportive.'' $\rightarrow$ Add a friendly opening sentence, keep the rest intact.
        \item Feedback: ``Add one concrete step.'' $\rightarrow$ Append one short sentence: ``You could start by setting a 5-minute timer and just begin.''
        \item Feedback: ``Organize into steps.'' $\rightarrow$ Turn a dense paragraph into 2--3 short bullets using the same ideas.
    \end{itemize}

    \textbf{FAIL-SAFE:}\\
    If none of the feedback is actionable (all are ``nothing/keep,'' or unsafe/contradictory), output the original comment unchanged (allow micro-polish only).\\

    \textbf{FINAL OUTPUT RULE:}\\
    Output only the revised comment, written in a natural human Reddit style. No analysis, no explanations, no formatting, no symbols. Avoid colons and em dashes in the final writing so it reads smoothly and conversationally.
    \endgroup
    \end{minipage}
}
\caption{\textbf{System prompt for Study 2, detailing instructions given to the LLM for applying multiple expert feedback to advice comments.}}
\Description{A boxed system-prompt template for Study 2 that instructs an LLM to minimally edit a Reddit-style advice comment using feedback from multiple expert reviewers. The figure lists \textbf{INPUTS} (post title, post body, original comment, and expert feedback notes separated by semicolons). A \textbf{CORE POLICY} section specifies that each reviewer note should be applied when possible, weighted equally, reconciled with minimal changes if conflicting, ignored if vague, and never used to introduce new recommendations beyond what reviewers request; it also prohibits inventing facts and constrains unsafe guidance. An \textbf{EDITING RULES} section requires minimal edits, preserving tone and phrasing, keeping length within $\pm$15\% of the original, and provides specific instructions for common requests (e.g., soften tone, add exactly one concrete step, convert to 2--3 bullets only if asked, replace numbers exactly as specified, remove suggestions without adding replacements unless provided). A \textbf{SAFETY \& BOUNDARIES} section prohibits medical or risky advice and directs adding at most a neutral boundary sentence if needed. An \textbf{OUTPUT} section requires returning only the revised comment text in natural Reddit style with no meta-commentary and constraints on punctuation and symbols. The figure also includes \textbf{TRANSFORMATION EXAMPLES} illustrating how to apply typical feedback, plus \textbf{FAIL-SAFE} and \textbf{FINAL OUTPUT RULE} instructions to output the original comment unchanged if feedback is non-actionable or unsafe.}
\label{fig:study2-intervention2}
\end{figure*}

\subsection{Study-2 Examples: Original Posts and Augmented Comments}
\label{appendix:study2_examples}

\subsubsection{``[Need Advice] How do I fix my life''}
\label{ex:fix-my-life}

\begin{quote}\small
I don't really have any friends. I have no job. The only thing going for me is that I'm not ugliest guy in the world and I'm in ok shape physically. My parents are also old but still alive and we talk occasionally, which I'm grateful for in a lot of ways, but I feel like once they are gone I'll have absolutely no support system. I have a bit of savings as well which is helping me get by but life feels confusing and difficult at times. I also feel old because I'm 31 and I don't have any parts of my life figured out at all.

I wish I felt like I was making a difference in people's lives even though I feel lost. Fitness is what I do currently but outside of that I don't have very many other positive habits or routines. At times I feel like I'm crazy because I can be somewhat miserable on more than one occasion and can't interact very well with other people. People are fed up for good reason. I don't have a life.
\end{quote}

\noindent\textbf{Reddit--Top (seed).}
\begin{quote}\small
if you're good at physical training and like to help people, learn to be a fitness instructor and help ppl that way
\end{quote}

\noindent\textbf{GPT-4o (seed).}
\begin{quote}\small
Hey there! First off, it's great that you're reaching out and looking to make positive changes. It sounds like you're in a tough spot, but the fact that you're in decent shape and have some savings is a solid foundation to build on. Here are a few steps you might consider:

1. \textbf{Set Small, Achievable Goals}: Start with something manageable. It could be as simple as waking up at the same time every day or dedicating 30 minutes to reading or learning something new. Small wins can build momentum.

2. \textbf{Explore New Interests}: Since fitness is already a part of your life, maybe look into related fields like nutrition or personal training. This could open up new career opportunities and help you meet like-minded people.

3. \textbf{Build a Routine}: Having a daily routine can provide structure and a sense of purpose. Include activities that promote well-being, like exercise, meditation, or journaling.

4. \textbf{Reconnect with People}: Try reaching out to old friends or family members. Even a simple text or call can help rebuild connections. You might also consider joining local groups or clubs related to your interests.

5. \textbf{Consider Professional Help}: If you're feeling stuck, talking to a therapist or counselor can provide guidance and support. They can help you work through past decisions and plan for the future.

Remember, it's never too late to make changes and improve your life. You're not alone in feeling lost, and taking small steps can lead to big changes over time. Keep pushing forward!
\end{quote}

\noindent\textbf{Human$\rightarrow$LLM-only.}
\begin{quote}\small
if you're good at physical training and like to help people, learn to be a fitness instructor and help ppl that way --- it gives you structure, purpose, and built-in social contact.

5-min starter: Google ``gyms hiring trainers [your city]'' and send 1 email, or DM a local trainer to shadow. Done.

If money/energy is tight:
\begin{itemize}
\item Get CPR/AED (cheap, 1 day). Many gyms hire floor staff and sponsor certs (ACE/NASM/ISSA).
\item Start small: coach one friend 1:1 for free, or run a free Saturday park workout for 1--3 friends/neighbors. Volunteer at Parkrun.
\item On low days: minimum = a 10-min walk + text 1 person.
\end{itemize}

Weekly feedback loop: did I move 3x, help 1 person, and make 1 outreach? If not, shrink the step and try again.

Also, 31 isn’t late --- and if depression/ADHD might be in the mix, a quick chat with a doc/therapist can make this way easier. If you’re in crisis: call/text 988 (US) or your local line.
\end{quote}

\noindent\textbf{Human$\rightarrow$LLM+Expert.}
\begin{quote}\small
Sorry you are going through this. Feeling lost, depressed, and isolated at 31 is brutal, and getting ghosted stings, but you are not a lost cause.

If you are good at training and like helping people, consider becoming a fitness instructor or getting a personal trainer license. It gives you structure, purpose, and a built-in community. Helping clients can pull your focus off the spiral in your head and onto making real progress with someone who wants it. You can only help people who want to be helped.

Also, try talking with a therapist or counselor you feel a connection with to work through the depression, understand yourself better, and start building support beyond your parents. You deserve help while you figure things out.
\end{quote}

\subsubsection{``[Need Advice] How are you supposed to do EVERYTHING?!''}
\label{ex:do-everything}

\begin{quote}\small
How are you supposed to study, and work, and spend time with your family, and learn a musical instrument, and clean the house, and cook, and read books, and watch movies, and buy groceries, and work out, and \dots

This is not a complaint or anything. I'm genuinely interested in how it is possible to live like this \dots{} like I really want to study, and I do want to work, and also I want to learn piano, and so on \dots{} but it just doesn't seem possible. Either there's not enough time, or I forget, or I'm too tired from one task and just want to go to sleep or \dots

So how does someone go about living like this? How can you do everything?

I'm 22, I've been trying to turn my life around for the past year. I have done many things I'm proud of since last year. I got way more productive than I ever was. But here's the thing: whatever I do, it doesn't seem like I can manage to do everything I want to \dots
\end{quote}

\noindent\textbf{Reddit--Top (seed).}
\begin{quote}\small
Random Advice: get a slow cooker and a rice cooker. You can go out to work or do stuff, come back and food is ready.

I still have no idea how people do so many things. Work takes the majority of my time but I still want to go back to school and do many other things too!
\end{quote}

\noindent\textbf{GPT-4o (seed).}
\begin{quote}\small
Hey there! It sounds like you're juggling a lot, and it's great that you're motivated to do so many things. The key here is to prioritize and manage your time effectively. Start by identifying your top priorities---what's most important to you right now? Focus on those and be okay with not doing everything at once.

It's also helpful to create a schedule or routine that includes time for rest and relaxation. Remember, it's not about doing everything perfectly, but about making consistent progress. You might also want to try habit stacking, where you pair a new habit with an existing one, to make it easier to incorporate into your day.

Lastly, don't forget to celebrate your achievements, no matter how small. You're already doing great by being more productive than before, so keep building on that!
\end{quote}

\noindent\textbf{Human$\rightarrow$LLM-only.}
\begin{quote}\small
Random Advice: get a slow cooker and a rice cooker so dinner is ready when you get home. It does not replace the joy of cooking, but it helps a little. Same here. Work takes most of my time and I still want to go back to school.

What helps me:
\begin{itemize}
\item Pick 2 priorities for this season (e.g., study + piano). Everything else = optional.
\item Make them tiny: 1 Pomodoro of study; 5 minutes of scales right after dinner (anchor it).
\item If you forget or are tired: leave the keyboard out, set a phone reminder, and have a 2-minute version on bad days.
\item Energy tiers: high day = 25 minutes study + 20 minutes workout; low day = read 2 pages + 5-minute walk or stretch.
\item Food: batch once, eat 3--4 times. Slow cooker + rice cooker = protein/veg in the morning; hit the rice when you get home.
\item Weekly 10-minute reset: what worked, what didn’t, tweak next week.
\end{itemize}

If you’re constantly wiped or forgetful despite decent sleep, consider a quick check-in with a doctor (iron/thyroid/ADHD, etc.).
\end{quote}

\noindent\textbf{Human$\rightarrow$LLM+Expert.}
\begin{quote}\small
Random advice: get a slow cooker and a rice cooker so dinner is ready when you get home. It does not replace the joy of cooking, but it helps a little. Same here. Work takes most of my time and I still want to go back to school.

Writing a short to-do list helps me arrange things and accept that I can't do everything at once. Pick what matters most this week, keep the list small, and give yourself credit for what you do get done. How do you all make time for everything?
\end{quote}

\end{document}